\documentstyle[aps,prb,twocolumn,epsfig]{revtex}

\newcommand{\sech}{\mathop{\mathrm sech}\nolimits}

\begin{document}

\preprint{PP DipFis UniCT 98-03}

\draft

\twocolumn[\hsize\textwidth\columnwidth\hsize\csname@twocolumnfalse%
\endcsname

\title{Sharp ${\bf k}$-space features in the order parameter within the  \\
       interlayer pair-tunneling mechanism of high-$T_c$ superconductivity}
\author{Giuseppe G.N. Angilella, Renato Pucci, Fabio Siringo}
\address{Dipartimento di Fisica dell'Universit\`a di Catania,\\
         and Istituto Nazionale per la Fisica della Materia,
         Unit\`a di Ricerca di Catania,\\
         57, Corso Italia, I-95129 Catania, Italy}
\author{Asle Sudb\o}
\address{Institutt for Fysikk, Norges Teknisk-Naturvitenskapelige
         Universitetet,\\
         Sem S{\ae}landsvei 9, Gl{\o}shaugen, N-7034 Trondheim, Norway}
\date{April 16, 1998; revised October 5, 1998}
\maketitle
\begin{abstract}
We study the ${\bf k}$-dependence of the gap function of a bilayer
   superconductor, using standard mean-field techniques applied to a 
   two-dimensional ($2D$) extended Hubbard model, in the presence of coherent 
   interlayer pair-tunneling and quenched coherent single-particle tunneling.
The \emph{intralayer} pairing potential thus defined is expandable in a 
   finite number ($5$) of basis functions for the irreducible representations 
   of the point-group of the perfectly square lattice, $C_{4v}$. 
This gives rise to a competition between  $s$- and $d$-wave symmetry, as the 
   chemical potential is increased from the bottom to the top of a realistic 
   band for most cuprates. 
It allows for  mixed-symmetry paired state at temperatures below $T_c$, 
   but never at $T_c$ on a square lattice.
Inclusion of the interlayer pair-tunneling into the effective pairing potential
   leads to highly non-trivial ${\bf k}$-space structures, such as pronounced 
   maxima along the Fermi line not seen in the absence of interlayer 
   pair-tunneling.
We show how such a gap structure evolves with temperature and with
   band filling, and how it affects various observables. 
In particular, a nonuniversal value of the normalized jump in the specific
   heat at $T_c$ will be evidenced, at variance with the conventional universal
   BCS result.
\end{abstract}

]


\section{Introduction}
\label{sec:intro}

The identification of the character of the asymptotic low-energy excitations
   of the the high-$T_c$ superconductors (HTCS) continues to present a 
   formidable 
   challenge to theorists and experimentalists in condensed matter physics.
These excitations are presumably a key feature in understanding the basics 
   of the phenomenon. \cite{Anderson:S92}
Although the superconducting state of the cuprates to a large extent in the 
   recent past has been considered conventional, it is becoming increasingly 
   clear that  such a statement requires certain modifications, to say the 
   least.~\cite{Poole:95a}
The latter statement is supported by recent experimental 
   findings.~\cite{Srikanth:97b,Srikanth:97a}

The controversy over the symmetry of the paired state ($s$- and
   extended $s$-wave vs higher order waves, particularly $d$-wave) and
   the coupling strength can nowadays be restated in more precise terms,
   due to the availability of samples with adequately pure composition
   and structure, and of improvements of experimental techniques.
   A central tool in this context is angle-resolved photoemission spectroscopy
   (ARPES),~\cite{Randeria:98a} with which one is able to extract, 
   if not the \emph{phase} of the superconducting order-parameter (OP),
   then at least the ${\bf k}$-dependence of its modulus
   at various temperatures and chemical compositions. 
Here, ${\bf k}$ is a wave-vector ranging over the first Brillouin zone 
   (1BZ) of the appropriate inverse lattice for the cuprate compound 
   under consideration.
In particular, there is a growing consensus on the occurrence of nodes of
   the OP along the $k_x = k_y$ direction in the 1BZ for optimally doped 
   Bi$_2$Sr$_2$CaCu$_2$O$_8$ (Bi2212).~\cite{Ding:95}
However, some contradictory claims ~\cite{Levi:93a} for 
   different samples seem to support, in a parallel way, the idea that 
   the detailed ${\bf k}$-space shape of the OP could be a material 
   specific property, although the location of the nodal lines may not 
   be.~\cite{Koltenbah:97a}
This is in agreement with the fact that both the critical temperature $T_c$
   itself and the maximum gap at $T=0$ change considerably from one material
   to another, as well as within a given material class, on varying the doping
   level. 

We will in this paper try to bring out a 
   few peculiar \emph{details} of some properties of the superconducting 
   state within the interlayer pair-tunneling mechanism (ILPT), 
   which seems to be almost unique to this pairing mechanism. 
It is at any rate becoming clear that the determination of the 
   location of nodal lines in the superconducting OP, i.e. its transformation 
   properties under the symmetry operations of the underlying lattice, by 
   no means suffices to unambiguously determine the unconventional pairing 
   mechanism. 
In this sense, the \emph{symmetry} of the OP is perhaps not a central issue, 
   although it certainly has been the focus of much research during the last 
   few years.
Moreover, the controversy over the symmetry of the OP  has initiated some of 
   the most sophisticated experiments in condensed matter physics 
   to date.~\cite{vanHarlingen:93,vanHarlingen:94,vanHarlingen:95,Kirtley:94} 

In this paper, we shall mainly consider the issue of gap anisotropy and 
   competition between different symmetry channels in the $2D$ extended 
   Hubbard model, characterized by a realistic band dispersion, including 
   nearest (N) and next-nearest (NN) neighbors hopping within the CuO planes, 
   and a small-range in-plane potential, allowing for in-site, N and NN 
   neighbors interaction, in the presence of pair tunneling 
   between adjacent layers.

The issue of the competition among symmetries in the gap function arising
   from the superconducting instability of an extended Hubbard model at
   a given band filling has previously been considered in the 
   literature,~\cite{Schneider:compact,Spathis:92,Li:93,O'Donovan:95}
   and has been recently addressed with renewed attention from both the
   theoretical and experimental points of view, in connection with the
   Cooper pair instability problem in lattice fermion systems,~\cite{Otnes:97a}
   and with the issue of material specific phenomenology in the 
   cuprates,~\cite{Koltenbah:97a} respectively.

The ILPT mechanism of high-$T_c$ superconductivity has been proposed as 
   a possible natural explanation for the observed high values of $T_c$ 
   in layered cuprates, as well as a number of other more difficult but
   related aspects of their complex 
   phenomenology.~\cite{WHA:87,Chakravarty:93a,Anderson:97}
On the other hand, neither the microscopic origin of the in-plane pairing
   nor the nature of the pairs has to be specified.
Several unconventional properties of these materials, due to strong 
   correlations already in the normal state, support the idea of a
   breakdown of Fermi liquid theory.
In particular, the absence of a Drude peak in the low frequency normal
   state $c$-axis optical conductivity, as observed in YBCO~\cite{Cooper:93} 
   and LSCO,~\cite{Uchida:96} would rule out metallic transport along the
   $c$-axis in the cuprates.
As a consequence, it has been suggested that \emph{coherent} single-particle
   interlayer tunneling is suppressed, due to the Anderson orthogonality 
   catastrophe,~\cite{Chakravarty:93a,Anderson:94,Clarke:94} whereas coherent
   pair tunneling in the superconducting phase is not restricted.

Among the mechanisms which would prevent single-particle tunneling,
   spin-charge separation~\cite{Anderson:95} has been proposed.
The tunneling process of one fermion would in fact require hopping of
   both spin and charge degrees of freedom, whereas a singlet object,
   like a Cooper pair, would carry charge $2e$ but no spin.

Therefore, coherent pair tunneling does not suffer from such restrictions, 
   and enters the total Hamiltonian as a second order effect in the single
   particle hopping matrix element, $t_\perp ({\bf k})$, whose dependence
   on the in-plane wave-vector ${\bf k}$ 
   (see Ref.~\onlinecite{Chakravarty:93a} and Eq.~(\ref{eq:tunn}) below) 
   has recently been confirmed by detailed band structure 
   calculations.~\cite{Andersen:96a}
The main aspect of the ILPT mechanism is that Josephson tunneling of Cooper 
   pairs between adjacent CuO layers dramatically amplifies the superconducting 
   pairing within each layer, by accessing the normal-state frustrated $c$-axis
   kinetic energy.

The addition of such a term to the total Hamiltonian does not only greatly 
   enhance $T_c$, but has also been able to describe the observed absence of the
   Hebel-Slichter coherence peak in NMR relaxation rate,~\cite{Sudboe:94a}
   as well as the recent neutron scattering experiments in optimally doped
   YBCO.~\cite{Yin:97a}
It was also recognized some time 
   ago~\cite{Sudboe:94b,Sudboe:95a,Yin:IJMP96} that, 
   in the same way as the ILPT mechanism very efficiently boosts the 
   magnitude of $T_c$ arising from the incipient pairing within the planes,  
   essentially due to its near ${\bf k}$-space diagonality, the amplitude as 
   well as the maximum value of the gap function are also dominated by the 
   effective coupling induced by the ILPT mechanism. \emph{Its actual  
   transformation properties under the symmetry operations of $C_{4v}$ are 
   however governed exclusively by the intralayer contribution to the pairing 
   kernel.}

In this paper, we shall make the latter statement more quantitative, showing
   how the interlayer coupling determines the detailed 
   ${\bf k}$-dependence of the gap, and actually tends to stabilize 
   one symmetry channel compared to other possible ones, as the chemical 
   potential is varied within the band.

This paper is organized as follows.
In Section~\ref{sec:model} 
   we introduce our model and review the basic formalism
   employed to derive the gap equations.
In Section~\ref{sec:gap} 
   we discuss the nontrivial numerical problems arising from
   the solution of the latter equations, due to the presence of the 
   ${\bf k}$-diagonal effective interlayer interaction.
A full discussion of the gap symmetry structure, its inherent anisotropy, its
   maximum values and locations thereof will be included. 
Calculations of the superconducting density of states (DOS) reveal remarkable
   structures, due to the gap anisotropy, which are believed to be 
   relevant to the observed anomalous phenomenology in tunneling
   junction characteristics.~\cite{Angilella:98a}
In Section~\ref{sec:temperature} 
   we shall address the issue of determining the critical
   temperature, as well as the temperature at which symmetry mixing occurs,
   as a function of the chemical potential.
At exactly $T=T_c$ the full ${\bf k}$-dependence of the gap function
   will be derived analytically, together with the critical exponents
   of the OP.
The resulting expression for the gap function in a closed form will serve
   as an evidence for the non-trivial anisotropic character and for the
   symmetry properties of the OP already at the critical point.
In Section~\ref{sec:applications} we shall
   consider various thermodynamical quantities in the superconducting phase.
Particular attention will be devoted to the normalized jump in the specific
   heat at the critical point, which, at variance with the BCS conventional
   result, turns out to be a nonuniversal number, due to symmetry competition
   and to the ILPT mechanism.
In Section~\ref{sec:summary} we summarize our results and present our 
   conclusions.

\section{The model}
\label{sec:model}

\subsection{Hamiltonian}

The model Hamiltonian we are going to consider in the following describes
   tightly bound interacting fermions in a bilayer complex:
\begin{equation}
   H = \sum_{{\bf k}\sigma i} \xi_{\bf k}^i 
         c_{{\bf k}\sigma}^{i\dag} c_{{\bf k}\sigma}^{i}
     + \sum_{{\bf kk^\prime}ij} \tilde{V}_{\bf kk^\prime}^{ij}
         c_{{\bf k}\uparrow}^{i\dag} c_{-{\bf k}\downarrow}^{i\dag}
         c_{-{\bf k^\prime}\downarrow}^{j} 
         c_{{\bf k^\prime}\uparrow}^{j} ,
\label{eq:Ham1}
\end{equation}
   where $c_{{\bf k}\sigma}^{i\dag}$ ($c_{{\bf k}\sigma}^{i}$)
   creates (destroys) a fermion on the layer $i$ ($i=1,2$), with 
   spin projection $\sigma$ along a specified direction, 
   wave-vector ${\bf k}$ belonging to the first Brillouin
   zone (1BZ) of a $2D$ square lattice, and band dispersion 
   $\xi_{\bf k}^i = \varepsilon_{\bf k}^i - \mu$, measured relative
   to the chemical potential $\mu$.
The second term in Eq.~(\ref{eq:Ham1}) describes an effective pair interaction,
   already restricted to the spin singlet channel only, with
\begin{equation}
   \tilde{V}_{\bf kk^\prime}^{ij} =
   \frac{1}{N} U_{\bf kk^\prime} \delta_{ij} 
   -T_J ({\bf k}) \delta_{\bf kk^\prime} (1-\delta_{ij} ),
\label{eq:pot1}
\end{equation}
   where $N$ is the number of sites in the square lattice, 
   $U_{\bf kk^\prime}$ measures the coupling interaction within each plane,
   and $T_J ({\bf k})$ is the tunneling matrix element between adjacent
   layers, motivated by Chakravarty {\em et al.}~\cite{Chakravarty:93a}
Equation~(\ref{eq:pot1}) shows, in particular, how the tunneling mechanism
   can be equivalently described by an interlayer effective interaction term,
   although \emph{local} in ${\bf k}$-space.

The main feature of this model is unusual. 
Although it can be cast in the form of a standard BCS-like
   effective Hamiltonian, the second term in the pairing
   potential arises from frustrated kinetic energy along the $c$-axis of the 
   cuprates, unaccessed in the normal state of the high-$T_c$ cuprates. 
However, it is \emph{lowered} on going into the superconducting state. 
This is a situation which has no counterpart in conventional Fermi-liquid 
   based superconductors, where the kinetic energy is \emph{enhanced} upon 
   going into the superconducting state, while being overcompensated by a 
   reduction in potential energy. 
In the above model superconductivity arises via a diametrically opposite 
   mechanism: Instead of having the gain in potential energy overcome 
   the loss of kinetic energy, it is the gain in kinetic energy that is 
   the driving mechanism.  There is ample experimental evidence that the 
   kinetic energy is lowered in the superconducting state of the cuprates. 
Although a confusing point has been that extracted values of the $c$-axis 
   penetration length has been consistent with estimates of Anderson based
   on gain in kinetic energy,~\cite{Anderson:98} they have also been consistent
   with $c$-axis conductivity sum rule arguments \emph{ignoring} the gain 
   in kinetic energy. 
This is traceable to subtleties in applications of $c$-axis conductivity 
   sum rules in unconventional metals, and a nice discussion clearing up this 
   crucial point has recently been presented by Chakravarty.~\cite{Chakravarty:98}

Comparison of band structure calculations~\cite{Yu:91a} with ARPES 
   results~\cite{Randeria:98a} for various cuprates suggest that the main
   hybridized single particle band crossing the Fermi level can be correctly
   described in the case of perfectly isotropic crystal symmetry
   by the tight-binding dispersion relation ($a$ being the lattice step)
\begin{equation}
   \varepsilon_{\bf k} = -2t [\cos(k_x a) + \cos (k_y a)] 
    + 4 t^\prime \cos (k_x a) \cos (k_y a) ,
\label{eq:disp}
\end{equation}
   where it has been recognized~\cite{Shen:93a} that at least nearest-neighbors
   ($t>0$) as well as next-nearest neighbors ($t^\prime >0$) hoppings 
   have to be
   retained, in order to reproduce the most relevant properties common to the
   mainly $2D$ band structure of the majority of the cuprate compounds.
First and foremost, we have in mind the \emph{shape} 
   of the Fermi surface, but also
   such features as the Van~Hove singularity in the density of states (DOS),
\begin{equation}
   n(\mu) = \frac{1}{2N} \sum_{\bf k} \delta (\varepsilon_{\bf k} -\mu)
\label{eq:nDOS}
\end{equation}
   at $\mu_{\mathrm VH} = -4t^\prime$, shifted towards the band bottom with 
   respect to the mid-band. 
We hasten to add that we are \emph{not} in any way implying that the Van~Hove 
   singularities in the single-particle density of states are important
   features in explaining the large critical temperatures in these 
   compounds.~\cite{Chakravarty:93a,Anderson:97,note:5}
General, and it seems to us very robust arguments for why the Van~Hove scenario
   is not viable, has been given by Anderson.~\cite{Anderson:97}
These considerations restrict $t^\prime /t \lesssim 0.5$, and imply
   a flat minimum at the 
   $\Gamma$ point, which gives rise to a pronounced, though finite, peak
   in the DOS at the band bottom. 
This band has a single-particle DOS which can be cast in closed form
   as~\cite{Otnes:97a,Xing:91}
\begin{equation}
   n(\varepsilon) = \frac{1}{2 \pi^2 t} \frac{1}{\sqrt{1- b \tilde\omega}}
   {\bf K} \left( \sqrt{\frac{1-[(b+\tilde\omega)/2]^2}{1-b \tilde\omega}} 
   \right),
\label{eq:DOS-Einar}
\end{equation}
   for $|(\tilde\omega+b)/2|<1$, and zero elsewhere.
In Eq.~(\ref{eq:DOS-Einar}) we have defined $b = - 2 t^\prime /t$, 
   $\tilde\omega=\varepsilon/(2t)$, and ${\bf K}(\alpha)$ is a complete 
   elliptic integral 
   of the first kind, with modulus $\alpha$.~\cite{Gradshteyn:80}
The DOS has a logarithmic singularity
   $n(\varepsilon) = (2\pi^2 t)^{-1} \sqrt{1-b^2}
   \left[ \log (8/(|\tilde\omega -b|)) + \log (\sqrt{1-b^2}) \right]$
   at $\varepsilon = 2 b t$, a finite cusp at the lower band-edge
   $n[\varepsilon=-2t(2+b)]=[4 \pi t (1+b)]^{-1}$, while the DOS
   at the upper band-edge is given by $n[\varepsilon=2t(2-b)]
   =[4 \pi t (1-b)]^{-1}$. 
These features of the DOS are important in stabilizing various 
   symmetry channels of the OP as the doping level is varied.~\cite{Otnes:97a}
A value of the nearest neighbors hopping parameter ranging around 
   $t=0.25$~eV satisfactorily models the band structure and the shape of 
   the Fermi surface of the majority of the cuprate 
   compounds.~\cite{Annett:90b,Norman:95a}
It is not among the main aims of this work to specify the 
   microscopic origin of the in-plane potential 
   $U_{\bf kk^\prime}$.~\cite{note:7}
However, any potential  with the symmetry of the underlying lattice
   may be expanded as a bilinear combination of basis functions for the
   irreducible representations of the crystal point group, which is $C_{4v}$
   for the $2D$ square lattice.~\cite{Annett:90a}
Assuming a finite-ranged potential, a finite subset of all the basis-functions
   (an infinite orthonormal set) will suffice.
Retaining therefore only on-site, nearest and next-nearest neighbors in-plane
   interactions, and projecting out interaction terms in the spin triplet 
   channel, one obtains
   the following expression for $U_{\bf kk^\prime}$, which
   is  \emph{separable} in ${\bf k}$-space:
\begin{equation}
   U_{\bf kk^\prime} = \sum_{\eta=0}^{4} \lambda_\eta 
   g_\eta ({\bf k}) g_\eta ({\bf k^\prime}),
\label{eq:pot2}
\end{equation}
where $g_0 ({\bf k} ) = 1$,
   $g_1 ({\bf k} ) = \frac{1}{2} [\cos (k_x a) + \cos(k_y a)]$,
   $g_2 ({\bf k} ) = \cos(k_x a) \cos(k_y a)$,
   $g_3 ({\bf k} ) = \frac{1}{2} [\cos (k_x a) - \cos(k_y a)]$,
   $g_4 ({\bf k} ) = \sin(k_x a) \sin(k_y a)$,
   and $\lambda_\eta$ ($\eta=0,1,\ldots 4$) are phenomenological effective
   coupling constants.
One immediately recognizes $g_0 ({\bf k} )$, $g_1 ({\bf k} )$, 
   $g_2 ({\bf k} )$ to display (extended) $s$-wave symmetry, whereas
   $g_3 ({\bf k} )$ and $g_4 ({\bf k} )$ display $d$-wave symmetry.
In the following, we shall assume repulsive on-site and attractive intersite
   coupling parameters ($\lambda_0 >0$ and $\lambda_1$, $\lambda_3 <0$),
   choosing their actual values in order to reproduce the correct order of
   magnitude for the critical temperature and gap maximum at $T=0$ for the
   cuprates. Throughout this paper, we keep $\lambda_2 = \lambda_4 = 0$.

Monte Carlo simulations support the idea that short-range antiferromagnetic
   fluctuations may produce an \emph{attractive} intersite 
   interaction (see Ref.~\onlinecite{Scalapino:95} for a review).
In our work, however, such an interaction is taken as \emph{phenomenological,}
   in the sense that an intersite attraction is at least required within
   an extended Hubbard model if one expects a $d$-wave contribution to the OP
   from the lowest lattice harmonics.
Remarkably, a perfectly tetragonal lattice requires $\lambda_1 = \lambda_3$.
Therefore, if one looks for $d$-wave coupling, i.e. a contribution from
   $g_3 ({\bf k})$ to $U_{\bf kk^\prime}$, then one should be also prepared to
   competition with extended $s$-wave contributions, coming at least from 
   $g_2 ({\bf k})$.

Finally, we assume the local dependence of the interlayer pair tunneling
   matrix element as $T_J  ({\bf k} ) = t_\perp^2 ({\bf k} )/t$,
   i.e. a second-order perturbation in the hopping matrix element 
   $t_\perp ({\bf k} )$ orthogonal to the CuO layers. 
Recent detailed band structure calculations~\cite{Andersen:96a} formally
   confirm the original choice of functional form made
   by  Chakravarty {\em et al.}:~\cite{Chakravarty:93a} 
\begin{equation}
   t_\perp  ({\bf k} ) = \frac{t_\perp}{4} [\cos(k_x a) - \cos(k_y a)]^2 ,
\label{eq:tunn}
\end{equation}
   which was arrived at by inspection of ARPES data combined with
   analyticity arguments.

In particular, ${\bf k}$-diagonality expresses conservation of the momentum
   component ${\bf k}_\parallel$ parallel to the CuO$_2$ plane during the
   hopping process.

We shall see in the numerical cases below that a fine tuning of $t_\perp$ 
   in the
   range $0.1$---$0.15$~eV is the main ingredient to reproduce the observed
   critical temperatures and zero-temperature gap maxima in different 
   compounds.
Such a range is however consistent with band structure calculations of 
   $t_\perp$.~\cite{Andersen:96a}

\subsection{Mean-field treatment}

A straightforward mean-field (MF) treatment of the total Hamiltonian
   Eq.~(\ref{eq:Ham1}) yields the approximate expression:~\cite{note:6}
\begin{equation}
   H_{\mathrm MF} = \sum_{{\bf k}\sigma i} \xi_{\bf k}^i 
         c_{{\bf k}\sigma}^{i\dag} c_{{\bf k}\sigma}^{i}
     + \sum_{{\bf k}i} [ \Delta^i_{\bf k}
         c_{{\bf k}\uparrow}^{i\dag} c_{-{\bf k}\downarrow}^{i\dag}
       + {\mathrm H.c.} ],
\label{eq:HamMF}
\end{equation}
where the auxiliary complex \emph{scalar} field (i.e., the gap function)
\begin{eqnarray}
   \Delta^i_{\bf k} &=& \sum_{j{\bf k^\prime}} 
   \tilde{V}_{{\bf kk^\prime}}^{ij} b_{\bf k}^j \nonumber\\
   &=&
   \frac{1}{N} \sum_{\bf k^\prime} U_{\bf kk^\prime}
   b_{\bf k^\prime}^i -T_J ({\bf k}) b_{\bf k}^{j} 
   (1-\delta_{ij}),
\label{eq:Delta1}
\end{eqnarray}
has been introduced.  
The gap function for the $i$th layer is thus seen to depend on the pair
   amplitude $b_{\bf k}^i = \langle c_{-{\bf k}\downarrow}^i
   c_{{\bf k}\uparrow}^i \rangle$ in the same layer, through the intralayer
   potential $U_{\bf kk^\prime}$, and on the pair amplitude in the adjacent
   layer, $b_{\bf k}^j$,
   through the interlayer tunneling amplitude $T_j ({\bf k})$,
   which acts as an effective potential, local in ${\bf k}$-space.

Equation~(\ref{eq:Delta1}) explicitly shows that, in general, the
   interlayer tunneling mechanism endows the gap function with a nontrivial,
   nonlocal structure in the direction orthogonal to the CuO layers.
Such a dependence is of course relevant in the more general case of 
   multi-layered compounds, and its consequences on $T_c$ have been studied,
   to some extent, by one of the present authors.~\cite{Sudboe:94b}
   A generalization of the methods of the present work to multi-layered
   systems, \emph{below} $T_c$ is straightforward, and is expected to unveil
   further features in the gap anisotropy, due to the coupling of the
   gap functions in adjacent layers.

In the case of a simple bilayer ($i=1,2$), the simplifying hypothesis that
   the pair amplitude $b_{\bf k}^i$ as well as the in-plane 
   single-particle band dispersion $\xi_{\bf k}^i$ and gap function
   $\Delta^i_{\bf k}$ do not depend on the layer index $i$ allows us
   to decouple the MF Hamiltonian Eq.~(\ref{eq:HamMF}) into a sum of
   independent Hamiltonians within each layer:~\cite{Chakravarty:93a}
\begin{equation}
   H_{\mathrm MF} = \sum_i \left\{ \sum_{{\bf k}\sigma} \xi_{\bf k} 
         c_{{\bf k}\sigma}^{i\dag} c_{{\bf k}\sigma}^{i}
     + \sum_{{\bf k}} [ \Delta_{\bf k}
         c_{{\bf k}\uparrow}^{i\dag} c_{-{\bf k}\downarrow}^{i\dag}
       + {\mathrm H.c.} ] \right\}.
\label{eq:HamMFdec}
\end{equation}
Standard diagonalization techniques in each layer then yield for 
   $\Delta_{\bf k}$ the BCS-like gap equation at the finite temperature
   $T$:
\begin{equation}
   \Delta_{\bf k} = - \sum_{\bf k^\prime} 
   \tilde{U}_{\bf kk^\prime} \chi_{\bf k^\prime} 
   \Delta_{\bf k^\prime},
\label{eq:gap1}
\end{equation}
where $\chi_{\bf k} = (2E_{\bf k} )^{-1} 
   \tanh (\beta E_{\bf k} /2 )$
   denotes the pair susceptibility, $\beta = (k_B T)^{-1}$, and 
   $E_{\bf k} = \sqrt{\xi_{\bf k}^2 + |\Delta_{\bf k} |^2}$ the
   upper band of the quasiparticle gapped spectrum.
In Eq.~(\ref{eq:gap1}) the pairing potential $\tilde{U}_{\bf kk^\prime}
   = \frac{1}{N} U_{\bf kk^\prime} - T_J ({\bf k}) 
   \delta_{\bf kk^\prime}$ includes the finite-range intralayer as well
   as the local interlayer effective interactions.
More explicitly, Equation~(\ref{eq:gap1}) reads
\begin{equation}
   \Delta_{\bf k} = - \frac{1}{1-T_J ({\bf k}) \chi_{\bf k}}
   \frac{1}{N} \sum_{\bf k^\prime} U_{\bf kk^\prime}
   \chi_{\bf k^\prime} \Delta_{\bf k^\prime}.
\label{eq:gap2}
\end{equation}
Making use of Eq.~(\ref{eq:pot1}) for the intralayer potential allows us to
   express the gap function as
\begin{equation}
   \Delta_{\bf k} = \frac{1}{1-T_J ({\bf k}) \chi_{\bf k}}
   \sum_\eta g_\eta ({\bf k}) \Delta_\eta ,
\label{eq:gap3}
\end{equation}
with
\begin{equation}
   \Delta_\eta = -\lambda_\eta \frac{1}{N} \sum_{\bf k^\prime}
   g_\eta ({\bf k^\prime}) \chi_{\bf k^\prime} 
   \Delta_{\bf k^\prime}.
\label{eq:gap4}
\end{equation}

At a generic temperature $T$, Equation~(\ref{eq:gap3}) does not yield
   immediately the explicit ${\bf k}$-dependence of $\Delta_{\bf k}$ 
   as it would in the limit of no interlayer tunneling ($T_J \to 0$).
This is due to the unusual prefactor 
   $[1-T_J ({\bf k}) \chi_{\bf k} ]^{-1}$, which includes
   $|\Delta_{\bf k} |$ self-consistently via the pair susceptibility
   $\chi_{\bf k}$.
However, this prefactor manifestly displays $s$-wave symmetry, since
   $|\Delta_{\bf k} |$ enters the pair susceptibility $\chi_{\bf k}$
   via the quasiparticle dispersion $E_{\bf k}$, which is an eigenvalue
   of $H_{\mathrm MF}$, and $T_J ({\bf k})$ has $s$-wave symmetry by
   itself [Eq.~(\ref{eq:tunn})].
Therefore, the complex parameters $\Delta_\eta$, which weigh the basis
   functions $g_\eta ({\bf k})$ in Eq.~(\ref{eq:gap3}), measure
   the contributions from the different symmetry channels to the full gap
   function, $\Delta_{\bf k}$, at a given temperature $T$.
We emphasize that such parameters are \emph{not} order parameters in
   themselves. 
Only $\Delta_{\bf k}$ as a whole serves as an OP for
   the superconductive instability, whose onset temperature $T_c$ is well
   defined and unique (see Sec.~\ref{sec:temperature} below).
However, a vanishing value of some of the $\Delta_\eta$ is a signal for the
   absence of the symmetry contribution which they represent to the full gap 
   function.
Besides, the set of parameters $\{ \Delta_\eta \}$ is not unique. 
Coefficients of any other complete orthonormal 
   set of functions would also suffice as a basis for expanding the gap 
   function.
The above choice of basis functions is convenient, since the 
   expansion of the in-plane pairing kernel used in this paper as bilinear 
   combination of basis functions truncates after just five terms
   ($\eta = 0,\ldots 4$).
However, irrespective of the choice of basis functions made, when all 
   contributions are summed up with appropriate weight-factors, the result 
   \emph{is} unique.
Moreover, the use of several parameters $\Delta_\eta$ does not attribute to 
   $\Delta_{\bf k}$ the structure of a multi-component (i.e., vectorial) 
   OP. We reserve the use of multi-component OP to situations encountered
   in systems such as $^3$He and possibly 
   UPt$_3$.~\cite{Fischer:89a,Sigrist:91}
The OP of  high-$T_c$ cuprates is much simpler, with only an amplitude 
   and a phase. 
On occasions, the set $\{ \Delta_{\eta} \}$ is referred to, incorrectly, 
   as the components of a multi-component OP of the cuprates. 

We finally remark that the self-consistent expression~(\ref{eq:gap3})
   endows $\Delta_{\bf k}$ with an inherently anisotropic 
   ${\bf k}$-dependence, which is modulated by the ${\bf k}$-harmonics
   $g_\eta ({\bf k})$, displaying explicit symmetry.
This is best seen in the limit $T\to T_c -0$, where the gap is vanishingly
   small, so that $\chi_{\bf k}$ in the right-hand side of 
   Eq.~(\ref{eq:gap4}) can be approximated by its normal state value,
   viz. $\chi_{\bf k}^{0c} = (2\xi_{\bf k} )^{-1} \tanh 
   (\beta_c \xi_{\bf k} /2 )$.
Only in such a limit, Equation~(\ref{eq:gap4}) already yields the explicit
   expression for the ${\bf k}$-dependence of the gap function, as a
   product of the anisotropic prefactor $[1-T_J ({\bf k}) 
   \chi^{0c}_{\bf k} ]^{-1}$ 
   and a superposition of ${\bf k}$-harmonics,
   weighted with vanishingly small coefficients $\Delta_\eta$.
We shall precise the latter statement in Sec.~\ref{ssec:anisTc} below, where
   the full ${\bf k}$-dependence of an incipiently opening gap function 
   at $T=T_c$ will be derived analytically.
It will be shown in Sec.~\ref{sec:temperature} that at $T=T_c$ the presence
   of a given symmetry contribution to the (just opening) gap function
   generally excludes mixing with other symmetries. The latter is 
   possible at lower temperatures, due to the highly nonlinear structure
   of the gap equations below the critical temperature, at least within a 
   given range of the band filling.
The mutual exclusion of orthogonal symmetries in the gap function at $T=T_c$
   is a well-known result in the case of nonlocal, separable 
   potentials.~\cite{Siringo:96a,Abrikosov:95}
We therefore
   recover this result also with ${\bf k}$-diagonal contributions to the 
   potential, such as the interlayer pair-tunneling effective interaction.

\section{Gap function anisotropy and symmetry}
\label{sec:gap}
\subsection{The auxiliary gap parameters $\Delta_{\eta}$}

Substituting $\Delta_\eta$ from Eq.~(\ref{eq:gap4}) into Eq.~(\ref{eq:gap3})
   yields
\begin{equation}
   \sum_{\eta^\prime} (\delta_{\eta\eta^\prime} + \lambda_\eta
   M_{\eta\eta^\prime} ) \Delta_{\eta^\prime} = 0,
\label{eq:gap5}
\end{equation}
with
\begin{equation}
   M_{\eta\eta^\prime} = \frac{1}{N} \sum_{\bf k} \tilde{\chi}_{\bf k}
   g_\eta ({\bf k}) g_{\eta^\prime} ({\bf k}),
\label{eq:matrices}
\end{equation}
where $\tilde{\chi}_{\bf k} = \chi_{\bf k} / [1-T_J ({\bf k})
   \chi_{\bf k} ]$ clearly acquires the role of a `renormalized' pair
   susceptibility.~\cite{Chakravarty:93a} 
These equations are in general coupled transcendental equations for
   $\Delta_{\bf k}$, and thus define a highly nonlinear problem.
Only at $T=T_c$ will the situation simplify considerably, as will be 
   discussed below.

However, once self-consistency has been achieved, Equations~(\ref{eq:gap5})
   are \emph{formally} linear and homogeneous in the \emph{phases} of the
   complex parameters $\Delta_\eta$, which are responsible for
   the overall complex phase of $\Delta_{\bf k}$, as shown by
   Eq.~(\ref{eq:gap3}).
Due to symmetry considerations, as remarked in Sec.~\ref{ssec:pure} below, 
   Equation~(\ref{eq:gap5}) reduces to two \emph{formally} independent sets
   of equations, with real coefficients, one for each group of parameters
   belonging to either $s$- or $d$-wave symmetry.
This means that the complex parameters $\Delta_\eta$ belonging to the same
   symmetry are all defined up to a \emph{same common} phase factor.
One can therefore speak of a relative phase between $s$- and $d$-wave 
   contributions.
In particular, it follows that there cannot be anisotropies in 
   ${\bf k}$-space of the \emph{phase} of the order parameter, other
   than the (trivial) one arising from eventual relative phase
   differences between two different symmetry contributions.
(This justifies the widely used terminology of $s+id$ symmetry, for example.)

Due to the presence of the unusual prefactor 
   $[1-T_J ({\bf k})\chi_{\bf k} ]^{-1}$ in Eq.~(\ref{eq:gap3}),
   which itself must be determined self-consistently by finding 
   $\Delta_{\bf k}$, 
   ordinary numerical procedures used to solve BCS-like gap equations in 
   the presence of separable 
   potentials~\cite{Otnes:97a,Spathis:92,Schneider:compact,ODonovan:compact} 
   are not applicable to the present case. 
Therefore, remarkably, the gap parameters $\Delta_\eta$ are not enough 
   to define the gap function completely: They yield information only 
   about its overall symmetry, on the degree of admixture of the various 
   symmetry channels in the gap function, and on their relative phase.
The solution for $\Delta_{\bf k}$, therefore, has to be obtained
   iteratively for each wave-vector ${\bf k}$ of interest.
The iterative numerical procedure employed to solve the gap equations
   is briefly outlined in App.~\ref{app:numsolv}.

We can  now 
   proceed with the solution of the gap equation  for each given value of 
   the chemical potential $\mu$ and temperature $T$.
We first keep $\mu$ at a fixed value.
By slowly decreasing the temperature from a relatively high value, we observe
   the appearance of a nontrivial solution to the gap equations,
   $\Delta_{\bf k}$, at a critical temperature $T_c = T_c (\mu)$, 
   whose value has been made comparable to the critical temperatures observed 
   in the cuprates, by a suitable tuning of parameters.
This onset is signalled by a nonvanishing value of \emph{some}
   of the parameters $\Delta_\eta$, corresponding to a nonzero contribution 
   of \emph{one} symmetry channel (Fig.~\ref{fig:detaT}).
We shall later show that only one orthogonal channel (restricting ourselves 
   in this work either to $s$- or to $d$-wave) can contribute to 
   $\Delta_{\bf k}$ at $T=T_c$ (see Sec.~\ref{sec:temperature} and 
   Ref.~\onlinecite{Siringo:96a}).
Upon further decreasing $T$ below $T_c$, $|\Delta_\eta (T)|$ increases
   (Fig.~\ref{fig:detaT}).
Together with $|\Delta_\eta |$, we plot in Fig.~\ref{fig:detaT} the maximum
   value of the gap function over the 1BZ,
\begin{equation}
   \Delta_M (\mu;T) = \max_{\bf k} |\Delta_{\bf k} (\mu;T)|.
\label{eq:dmaxmu}
\end{equation}

One immediately recognizes that $\Delta_M$ is considerably enhanced with
   respect to $|\Delta_\eta |$, which are representative of the values it
   would have had, in the absence of ILPT.

We shall later show analytically (see Sec.~\ref{ssec:anisTc} below) that
   $\Delta_\eta$ and $\Delta_M$ behave like
   $\sim (T_c -T)^{1/2}$ at $T=T_c -0$, as it is expected in any
   mean-field theory for an OP.
However, Figure~\ref{fig:detaT} shows that the behavior of some $\Delta_\eta$
   as functions of $T$ may soon depart from its critical limit close to $T_c$,
   depending on the value of the chemical potential.
This is to be contrasted with the dependence on temperature of $\Delta_M$, 
   which closely resembles the conventional one in BCS theory.
The unconventional temperature dependence of the parameters $\Delta_\eta$
   below $T_c$ directly stems from their definition, Eq.~(\ref{eq:gap4}).
A different choice of parameters $\{ \Delta_\eta \}$ would in general lead
   to a different temperature dependence, except their critical behavior
   at $T_c$.
On the contrary, we expect the result obtained for $\Delta_M$ to be unique,
   as its value depends more on the ILPT amplitude than on the parametrization
   employed for the symmetry character of the OP.

Depending on the value of $\mu$, other symmetry channels may begin contributing
   to the full order parameter $\Delta_{\bf k}$ as $T$ decreases.
This is signalled by a nonzero value of the remaining parameters $\Delta_\eta$,
   and by an enhancement of the parameters $\Delta_\eta$ corresponding to the
   symmetry channel already active, as in the numerical example shown in
   Fig.~\ref{fig:detaT}.
The critical exponent with which the new $\Delta_\eta$'s open at the critical
   temperature is again $1/2$, as can be shown 
   analytically~\cite{Angilella:thesis}.
The temperature $T_m = T_m (\mu)$ at which this happens does \emph{not}
   correspond to any new instability: The system is already a superconductor,
   with massive gauge-fluctuations and a finite superconducting coherence 
   length.
No remarkable feature is to be observed in $\Delta_M$ as a function of $T$. 
Its value depends more on the anisotropy induced by the interlayer tunneling 
   mechanism than on the intralayer potential.
At the mean-field level, the OP enhances its overall amplitude and its 
   anisotropy character, by allowing pairs to condense in more symmetry 
   channels.

Symmetry mixing is made possible by the nonlinear character of the gap 
   equations themselves, which becomes increasingly  more relevant as
   the temperature decreases towards $T=0$, given our choice of an extended
   in-plane real-space pairing potential.
Such a possibility has been already studied in detail by 
   Spathis {\em et al.,}~\cite{Spathis:92} who used a description in 
   terms of a bifurcation of the gap parameters, and by 
   O'Donovan and Carbotte~\cite{ODonovan:compact} 
   in the case of an extended Hubbard model without interlayer pair-tunneling.
Consistent results have also been obtained by Otnes and one of the present
   authors~\cite{Otnes:97a} for the Cooper problem in presence of an
   extended intralayer Hubbard potential.

Inclusion of a ${\bf k}$-diagonal interlayer pair-tunneling term in such 
   a model preserves this  feature. 
A novel effect of the interlayer
   pair-tunneling is that it strongly influences the competition between 
   $s$- and $d$-wave symmetry channels in the OP, enhancing 
   a dominant symmetry channel compared with the subdominant other one. 
The matrix element $T_J({\bf k})$ generally reduces the region of 
   symmetry mixing in the $(\mu,T)$ phase diagram, as will be discussed
   more in detail  in Sec.~\ref{sec:temperature}. 
The reason is 
   that when a gap amplitude starts to grow at $T=T_c$,  the dominant
   channel will initially suppress pairing in other channels.
This is
   generic to any superconductor allowing mixed symmetries to appear
   in th OP, also conventional ones. 
Furthermore, it is important to note that 
   the gap at a certain ${\bf k}$-point in the BZ depends on the gap at 
   all other ${\bf k}$-points  via the nonlocality of the intralayer part 
   of the pairing kernel, even though the inter-layer part is local. 
The result is a strong enhancement
   of the gap amplitudes in the dominant channels, which then to an even 
   stronger degree will suppress competing channels. 
The consequences of a possible symmetry mixing on observables will be analyzed 
   below in Sec.~\ref{sec:applications}. 
We note, however, that due to the 
   specific choice of band structure and intra-plane coupling constants, it is 
   well established that $d$-wave pairing will dominate in the vicinity of 
   half-filling, while $s$-wave pairing wins out for low filling fractions. 
Hence, in the cuprates, $T_J$ will tend to \emph{stabilize} $d$-wave pairing
   compared to competing channels, were $d$-wave pairing to be the dominant 
   intralayer channel.

\subsection{The order parameter $|\Delta_{{\bf k}}|$}

Primarily, the interlayer tunneling amplitude $T_J ({\bf k})$ in 
   Eq.~(\ref{eq:gap2}) affects the overall anisotropic structure of the gap
   function, and not its symmetry character.
To show this, the dependence of $\Delta_\eta$ on $T$ at a given chemical 
   potential $\mu$ does not suffice alone.
Therefore, in Fig.~\ref{fig:dkT} we show the overall ${\bf k}$-dependence
   of $|\Delta_{\bf k} |$ over the whole 1BZ
   at $T=0$, for a fixed value of $\mu$.
We choose to plot $|\Delta_{\bf k} |$ along the family of mutually
   orthogonal lines defined by $\varepsilon_{\bf k} = {\mathrm const}$
   and $\gamma_{\bf k} = {\mathrm const}$, where $\gamma_{\bf k}$
   is a harmonic conjugate of $\varepsilon_{\bf k}$.
Such a choice is best suited to exhibit and highlight the structure of maxima 
   in the gap function along $\xi_{\bf k} =0$.

From the numerical analysis,
   one clearly observes a nodal line along the $k_x = k_y$ direction for
   $T_m < T < T_c$, which evolves into a line of local minima as symmetries
   mix below $T_m$ down to $T=0$.
Moreover, what is more apparent is the presence of rather pronounced lines
   of maxima whose location in the 1BZ follow the locus of the dispersionless
   wave-vectors for the normal state quasiparticles, i.e. the would-be Fermi
   line, defined by $\xi_{\bf k} =0$.
Absolute maxima (sharp peaks) are located at the intersection of the 
   $\xi_{\bf k} =0$ locus with $k_y =0$ for $\mu < \mu_{\mathrm VH}$ 
   (corresponding to a 
   Fermi line closed around the $\Gamma$ point), or with $k_x =\pi/a$
   for $\mu > \mu_{\mathrm VH}$ (corresponding to a Fermi line closed 
   around $M=(\pi/a,\pi/a)$).
Such features are of course produced by the enhancing prefactor
   $[1-T_J ({\bf k})\chi_{\bf k} ]^{-1}$ 
   in Eq.~(\ref{eq:gap2}), which gives its maximum contribution where
   $T_J ({\bf k})\chi_{\bf k} \approx 1$, i.e. exactly as quoted
   above.~\cite{note:2}
The reason for the \emph{sharpness} of these features is the 
   ${\bf k}$-diagonality of the inter-layer pair-tunneling term. 
Similar spikes are difficult to obtain 
   with more conventional, i.e. ${\bf k}$-nondiagonal, contributions to the
   pairing kernel;~\cite{Perali:96} 
   in such cases we do not get the unusual enhancement factor 
   $[1-T_J ({\bf k}) \chi_{\bf k}]^{-1}$ in the effective pairing 
   susceptibility responsible for the peaks, and anisotropies in the 
   pairing kernel tend to be smeared by integrations. 
We suggest that improved energy resolution in ARPES 
   is a useful tool to look for sharp features in the gap on the 
   Fermi surface, which appears to be a hallmark of the ILPT mechanism.

The maxima distribution and values of $\Delta_{\bf k}$ along the
   Fermi line is in qualitative and quantitative agreement with
   high-resolution photoemission data available for the bilayer
   Bi2212.~\cite{Ding:95}
It is gratifying to recover such results, without making detailed
   reference to the bilayer band structure.~\cite{Temmermann:96} 
It requires invoking the ILTP mechanism, 
   where the amplitude $T_J ({\bf k})$ 
   depends on ${\bf k}$ through Eq.~(\ref{eq:tunn}) 
   (Ref.~\onlinecite{Chakravarty:93a})
   in a way which is confirmed by band structure 
   calculations.~\cite{Andersen:96a}

Together with a remarkable ${\bf k}$-dependence of the order parameter,
   one observes a different temperature variation of $\Delta_{\bf k}$
   depending on the location of ${\bf k}$ in the 1BZ, and particularly
   along the Fermi line $\xi_{\bf k} =0$, where anisotropy is enhanced.
This is in qualitative agreement with recent ARPES measurements of 
   $\Delta_{\bf k}$ in underdoped Bi2212 at different points of the Fermi
   line.~\cite{Norman:98a}

\subsection{Superconducting DOS}

One consequence of such a peculiar anisotropy is e.g.
   given by the superconductive 
   density of states at $T=0$,
\begin{equation}
   n_S (\omega) = \frac{1}{N} \sum_{\bf k} 
   [u^2_{\bf k} \delta (\omega-E_{\bf k} ) + 
    v^2_{\bf k} \delta (\omega+E_{\bf k} )],
\label{eq:sDOS}
\end{equation}
   where
\begin{mathletters}
\begin{eqnarray}
   u^2_{\bf k} &&= \frac{1}{2} \left( 1 + 
                       \frac{\xi_{\bf k}}{E_{\bf k}} \right),\\
   v^2_{\bf k} &&= \frac{1}{2} \left( 1 -
                       \frac{\xi_{\bf k}}{E_{\bf k}} \right)
\end{eqnarray}
\end{mathletters}
are the usual expressions for the coherence factors in BCS-like 
   theories,~\cite{Enz:92} which hold for an interacting Fermi liquid,
   in the absence of spectral anomalies.~\cite{Yin:IJMP96}
Equation~(\ref{eq:sDOS})
   obviously reduces to $n(\omega)$, Eq.~(\ref{eq:nDOS}), in the limit
   $\Delta_{\bf k} \to 0$.
In Fig.~\ref{fig:sDOS}, we plot the superconducting DOS $n_S (\omega)$ 
   corresponding to a superconducting spectrum
   $E_{\bf k}$ with fully anisotropic, prevalently $d$-wave gap function
   $\Delta_{\bf k}$, obtained self-consistently at $T=0$, and the analogous
   quantity $n_S^d (\omega)$, where a pure $d$-wave gap function 
   $\Delta^d_{\bf k} = \Delta^d g_3 ({\bf k})$ has been used,
   with $\Delta^d = \max_{\bf k} |\Delta_{\bf k} |$.

In both cases, a gap opens in the SC spectrum at $\omega=0$ (i.e., around
   the Fermi level).
However, the minimum at $\omega=0$ in $n_S^d (\omega)$ is flatter than in
   $n_S (\omega)$, and the features around $\omega =0$ are
   quite less pronounced and less asymmetric with respect to the Fermi level.
Such behavior in the superconducting DOS is peculiar to the interlayer
   tunneling mechanism, and is promising~\cite{Angilella:98a} in order to
   explain the anomalous features observed in tunneling junctions experiments
   with Bi2212.~\cite{Wei:98}

To complete our picture of the competition of gap symmetries and anisotropy
   in the ILPT mechanism, we evaluated $\Delta_{\bf k}$
   at $T=0$ for chemical potential $\mu$ ranging from the bottom to the top
   of the band.
In Fig.~\ref{fig:detamu}, we plot $|\Delta_\eta (\mu;T=0)|$ against $\mu$.
One observes that $s$-wave symmetry prevails at low band filling, and $d$-wave
   symmetry at higher filling, which is consistent with earlier 
   results.~\cite{Spathis:92,Schneider:compact,ODonovan:compact,Otnes:97a}
In a rather narrow region, an OP with mixed symmetry occurs.
Numerical analysis revealed that the ILPT mechanism reduces
   the extension of the latter
   with respect to the limit $T_J \to0$, thus showing that
   a local non-separable contribution to the pairing potential frustrates,
   in general, the coexistence of orthogonal symmetries at low temperatures.
We argue, therefore, that a true, generally non-separable potential, of
   which Eq.~(\ref{eq:pot2}) is only a truncated expansion over a reduced
   set of basis functions, could even suppress symmetry mixing entirely.
We have however no formal proof of a such a statement, at present.

From a numerical analysis of the gap maximum at $T=0$, $\Delta_M^0 (\mu) =
   \Delta_M (\mu,T=0)$ [Eq.~(\ref{eq:dmaxmu})],
   as a function of $\mu$ (Fig.~\ref{fig:detamu}), we moreover conclude 
   that the ILPT mechanism
   yields reasonably large values of the gap maximum, as observed
   experimentally in the HTCS,~\cite{Chakravarty:93a} and that the actual
   values of the intralayer coupling constants $\lambda_\eta$ contribute
   only in a minor way.
Furthermore, Fig.~\ref{fig:detamu} shows that the largest gaps correspond to
   prevalently $d$-wave symmetry, and are obtained for 
   $\mu \approx \mu_{\mathrm VH}$ (the exact location depending weakly on 
   $\lambda_\eta$), where the enhancement due to $T_J ({\bf k})$ is
   highest, once more showing the relevance of the $2D$
   character of the single particle dynamics in the normal state through
   their dispersion relation, and the importance of the actual value of
   the next-nearest neighbors hopping amplitude $t^\prime$ in 
   Eq.~(\ref{eq:disp}), which fixes the value of $\mu$ at which the Fermi
   line changes its topology.

\section{Critical temperature}
\label{sec:temperature}

Among the many experimental facts concerning the HTCS phenomenology
   that the ILPT mechanism is able to describe, probably
   the most apparent is the ease with which the high value of the critical 
   temperature itself is explained. 
This is first and foremost due to
   the ${\bf k}$-diagonality of the intralayer part of the kernel, and
   has previously been investigated in some detail by Chakravarty 
   {\em et al.},~\cite{Chakravarty:93a} when considering the ILPT 
   mechanism for bilayer compounds such as Bi2212. 
Of course, it is
   a matter of some importance to investigate the effect of inelastic
   scattering, i.e. ${\bf k}$-space broadening, of the interlayer term,
   to investigate how detrimental effect it has on $T_c$. 
Preliminary 
   results \cite{Fjaerestad:98} show that $T_c$ is fairly robust to a 
   broadening of the interlayer term.

In this Section, we generalize the results of Ref.~\onlinecite{Chakravarty:93a}
   to arbitrary doping, conveniently
   reparametrized by the chemical potential ranging within the dispersion
   bandwidth, extending the analysis to the case of the intralayer potential 
   proposed in Eq.~(\ref{eq:pot2}).
The dependence of $T_c$ on $\mu$ is a relevant point in itself, since it
   allows to clarify the role of the $2D$  hole dynamics and that of the 
   incoherent, interlayer pair-tunneling mechanism in determining 
   the shape and 
   extension of the $(\mu,T)$ region allowed for the superconductive 
   instability to occur.

A separate question, in the present context, concerns the $(\mu,T)$ region
   allowed to superconductivity characterized by a symmetry order parameter.
Due to the structure of the gap equation~(\ref{eq:gap2}), such a question
   involves considerable numerical difficulties, in comparison with previous
   work of some of the present authors,~\cite{Angilella:96b} which will be
   dealt with in some detail.

\subsection{Superconducting instability: pure symmetry}
\label{ssec:pure}

At $T=T_c$, the mean-field gap function $\Delta_{\bf k}$ is vanishingly
   small everywhere in the ${\bf k}$-space.
Therefore, Eq.~(\ref{eq:gap5}) linearizes to
\begin{equation}
   \sum_{\eta^\prime} (\delta_{\eta\eta^\prime} + \lambda_\eta 
   M_{\eta\eta^\prime}^0 ) \Delta_{\eta^\prime} = 0,
\label{eq:gap5lin}
\end{equation}
   where the linearized matrix elements
\begin{eqnarray}
   M_{\eta\eta^\prime}^0 &=& \lim_{\Delta_{\bf k}\to 0} M_{\eta\eta^\prime} 
   \nonumber\\
   &&=
   \frac{1}{N} \sum_{\bf k} \frac{\chi^0_{\bf k}}{1-T_J 
   ({\bf k}) \chi^0_{\bf k}} g_\eta ({\bf k}) g_{\eta^\prime}
   ({\bf k})
\label{eq:matriceslin}
\end{eqnarray}
   do not depend on $\Delta_{\bf k}$ any more. 
These matrix elements are analogs of the well-known 
   logarithmically divergent integrated  pairing susceptibility in the 
   BCS-theory.~\cite{Agd:63}
Here, what appears are
   integrated, effective pairing susceptibilities, projected down on various
   symmetry channels. Symmetry dictates that only basis functions having
   the same transformation properties, albeit belonging to different 
   irreducible representations of $C_{4v}$, can yield a finite effective 
   pairing susceptibility  $M_{\eta \eta^\prime}^0$. 

The condition for 
   Eq.~(\ref{eq:gap5lin}) to have a nontrivial solution 
   $\{\Delta_{\eta^\prime} \}$ is that
\begin{equation}
   \det (\delta_{\eta\eta^\prime} + \lambda_\eta M_{\eta\eta^\prime}^0 ) = 0.
\label{eq:nontrivial1}
\end{equation}
Due to the $s$-wave symmetry character of $\tilde{\chi}^0_{\bf k}$ and the 
   definite symmetry character of the basis functions $g_\eta ({\bf k})$, 
   $M_{\eta\eta^\prime}^0$ is block-diagonal. 
Its elements are 
   nonzero if and only if $\eta$ and $\eta^\prime$ denote symmetry channels
   belonging to the same irreducible representation of the crystal point group.
Therefore, Eq.~(\ref{eq:nontrivial1}) for $T_c$ at a given $\mu$ factorizes
   into
\begin{equation}
   D_{\lambda_s}^0 (\mu,T) D_{\lambda_d}^0 (\mu,T) = 0,
   ~~~~~~ T=T_c.
\label{eq:sd}
\end{equation}
   Here, $D_{\lambda_h}^0 (\mu,T_c ) = \det ( \delta_{\eta_h \eta_h^\prime}
   + \lambda_{\eta_h} M^0_{\eta_h \eta_h^\prime} ) ~~(h=s,d)$ depends
   only on a subset of the $\lambda_\eta$ ($\eta =\eta_h$).
Linearization therefore decouples the two symmetries at $T=T_c$.
The solution correponding to the largest value of $T_c$ from Eq.~(\ref{eq:sd}) 
   corresponds to the true superconducting transition temperature. 
The transformation properties of the corresponding eigenvectors determine 
   in which (single) symmetry channel the dominant superconducting instability
   occurs. 
At $T=T_c$, the other solution correponds to a subdominant superconducting 
   instability, and is physically irrelevant. 
Generically, precisely at $T=T_c$,  we thus cannot have an instability into 
   a mixed state, i.e. a superconducting instability with 
   eigenvectors having components belonging to different irreducible 
   representations of $C_{4v}$. 
A mixing of symmetries can only occur below the physical $T_c$, 
   as discussed more in detail in Sec.~\ref{sec:gap}.
The exception to this statement
   occurs when $\mu$ is fine-tuned such that the zeroes of the $d$- and 
   $s$-determinants are found at the same temperature. 
The phase space for this to occur is however vanishingly small.
Such a result is a generalization of a known theorem, which applies to
   purely nonlocal separable extended 
   potentials.~\cite{Siringo:96a,Abrikosov:95}
The generalization has been made possible by the definite symmetry character
   ($s$-wave) of the effective local potential induced by the interlayer
   tunneling amplitude $T_J ({\bf k})$ in Eq.~(\ref{eq:pot1}),
   and is of course extendible to potentials supporting an arbitrary number of
   symmetry channels in the OP.~\cite{Fehrenbacher:95}
We emphasize that these statements pertain to the square lattice only. 
In systems with pronounced $ab$-plane orthorhombicity, such as YBCO,
   a certain amount of mixing  is expected on quite general grounds,
   and is indeed inevitable. 
The underlying lattice point group is $C_{2v}$, and thus an expansion in 
   terms of basis functions for $C_{4v}$ will yield several terms. 
   
The issue of determining $T_c = T_c (\mu)$ and the symmetry channel in which 
   the instability occurs, proves therefore to be equivalent to comparing 
   the two solutions of Eq.~(\ref{eq:sd}). 
It must be noted, however, that at variance with the case of no interlayer
   tunneling, the linearized matrix elements $M^0_{\eta\eta^\prime}$
   included in the definitions of $D_{\lambda_h}^0 (\mu,T_c )$ display a
   divergent behavior at some value $T_c = T^\star (\mu)$, due to the
   presence of the denominator $1-T_J ({\bf k})\chi^0_{\bf k}$ in the
   summand of Eq.~(\ref{eq:matriceslin}).
It has already been emphasized~\cite{note:2} that the self-consistency 
   condition Eq.~(\ref{eq:gap2}) for a nonvanishing gap $\Delta_{\bf k}$ 
   below the true $T_c$ prevents the occurrence of such a singularity. 
The singularity  is due only to the mathematical artifact
   of extending the definitions of the determinants  
   $D_{\lambda_h}^0 (\mu,T_c )$ to a domain below their zeroes. 
This is not physically meaningful, since
   the opening of $\Delta_{\bf k}$ modifies  their very definitions.

The occurrence of such an unusual singularity in the integrated  effective 
   pairing-susceptibility, projected on various symmetry channels,
   has a physical meaning. 
It shows how the action of an interlayer pair-tunneling mechanism bounds 
   the critical temperature from below, and therefore enhances it.
Given an intraplane contribution to the pairing-kernel, a lower bound
   on $T_c$ is set by the matrix element $T_J$; the lower bound
   roughly given by $T_J/4$ (see Eq.~(\ref{eq:Tstar}) below).
Beyond this, the actual value of $T_c$ is fixed by the intralayer coupling 
   symmetry and strength.
As already noted, $\chi^0_{\bf k}$ is maximum along the Fermi line,
   where $\lim_{T\to T_c} \lim_{\xi_k \to 0} \chi^0_{\bf k} = \beta_c /4$.
Therefore, the renormalized susceptibility $\tilde{\chi}^0_{\bf k}$ along
   the Fermi line is maximum where $T_J ({\bf k})$ is maximum, i.e.
   at the intersection of the Fermi line with the $\Gamma$--$X$--$M$ path
   in the 1BZ (and symmetry related points).
Looking for the highest temperature $T^\star (\mu)$ at which the maximum
   of $\tilde{\chi}^0_{\bf k}$ diverges, one has therefore to distinguish
   between the two possible topologies for the Fermi line arising from
   Eq.~(\ref{eq:disp}).
One finds, analytically,
\begin{equation}
   k_B T^\star (\mu) = \left\{
      \begin{array}{ll}
         \frac{T_J}{64} \left( \frac{\mu_\bot -\mu}{\mu_\bot +2t} \right)^4 ,
       & \mu_\bot \leq \mu < \mu_{\mathrm VH} ,\\
         \frac{T_J}{64} \left( \frac{\mu_\top -\mu}{\mu_\top -2t} \right)^4 ,
       & \mu_{\mathrm VH} \leq \mu \leq \mu_\top  ,
      \end{array} \right.
\label{eq:Tstar}
\end{equation}
   where $\mu_\bot$, $\mu_\top$ denote the bottom and the top of the band,
   respectively, which generalizes the expression given in 
   Ref.~\onlinecite{Chakravarty:93a}.
At the Van~Hove singularity, $T^\star (\mu)$ is maximum, with 
   $k_B T^\star (\mu_{\mathrm VH} ) = T_J /4 = t_\bot^2 /4t$,
   yielding a lower bound $k_B T_c \lesssim 0.01$~eV
   ($T_c \lesssim 110$~K), which is a representative value for most
   bilayer cuprates.

Figure~\ref{fig:tczero} shows our results for $T^\star$ and $T_c$
   as functions of $\mu$.
The values of the parameters have been chosen as quoted in order to yield
   a critical temperature at optimal doping whose value is representative
   of the bilayer cuprate superconductors.
Superconductivity appears restricted predominantly to the lower part of the 
   band, even though a nonvanishing lower bound $T^\star (\mu)$ assures
   a nonzero, albeit decreasing, $T_c$, as $\mu$ increases towards
   the top of the band.
In that regime, however, we showed numerically that $\Delta_M^0$ is vanishingly
   small (cf. Fig.~\ref{fig:detamu}).
As previously observed,~\cite{Angilella:96b,Otnes:97a} $s$-wave symmetry
   prevails near the bottom of the band, whereas $d$-wave symmetry wins out as
   $\mu$ increases. 
A robust qualitative argument for this was given in 
   Ref.~\onlinecite{Otnes:97a}.
The critical temperature $T_c$ reaches its optimal value near 
   $\mu=\mu_{\mathrm VH}$, the exact location depending  on the
   set of values $\{ \lambda_\eta \}$ actually chosen for the intralayer
   coupling parameters.

\subsection{Gap anisotropy at the critical point}
\label{ssec:anisTc}

The ILPT mechanism is seen to strongly enhance the ${\bf k}$-space anisotropy
   of the gap function also at $T=T_c$, regardless of the symmetry character
   that the OP takes on, which at the critical point is unambiguosly defined
   (no mixing).
This is already apparent from Eq.~(\ref{eq:gap3}), and can be proved by
   exhibiting the full analytical ${\bf k}$-dependence of the gap function
   $\Delta_{\bf k}$, at $T=T_c -0$.

For $T \lesssim T_c$, one may Taylor expand all quantities of interest in
   powers of $\beta^2 |\Delta_{\bf k} |^2 
   \ll 1$, safely retaining the first nonzero term only. 
From Eq.~(\ref{eq:matrices}), one obtains:
\begin{eqnarray}
   M_{\eta\eta^\prime} &=& M_{\eta\eta^\prime}^0 
                         - \beta^3 \frac{1}{N} \sum_{\bf k}
        \frac{\phi(\beta\xi_{\bf k} /2)}{[ 1 -T_J ({\bf k}) \chi^0_{\bf k} ]^2}
        g_\eta ({\bf k}) g_{\eta^\prime} ({\bf k})
        |\Delta_{\bf k} |^2 \nonumber \\ 
                         &&+ {\cal O} (\beta^2 |\Delta_{\bf k} |^2 ),
\label{eq:matrices-expansion}
\end{eqnarray}
   where
\begin{equation}
   \phi(x) = \frac{1}{32 x^3} \left(\tanh x - x \sech^2 x \right),
\label{eq:aux-f}
\end{equation}
   and a superscript zero denotes that the limit $\Delta_{\bf k}  \to 0$
   has been taken.
At $T=T_c$, only one symmetry channel is active, therefore 
   Equation~(\ref{eq:sd}) is satisfied by the vanishing of one block 
   determinant, say $D_{\lambda_h}^0 (\mu, T_c ) = 0$,
   ($h = s$ or $d$).
Expanding $D_{\lambda_h} (\mu,T)$ around $\beta^2 |\Delta_{\bf k} |^2 =0$ and
   making use of Eq.~(\ref{eq:matrices-expansion}), one finds:
\begin{eqnarray}
   D_{\lambda_h} (\mu,T) &=& D_{\lambda_h}^0 (\mu,T) - \beta^3 \frac{1}{N}
   \sum_{\bf k}
        \frac{\phi(\beta\xi_{\bf k} /2)}{[ 1 -T_J ({\bf k}) \chi^0_{\bf k} ]^2}
   \nonumber\\
   &&{}\times \left( 
        {\sum_{\eta\eta^\prime}}^h 
        \lambda_\eta g_\eta ({\bf k}) W_{\eta\eta^\prime}^0 
                     g_{\eta^\prime} ({\bf k})
   \right)
        |\Delta_{\bf k} |^2 \nonumber\\
   &&\phantom{{}\times(} + {\cal O} (\beta^2 |\Delta_{\bf k} |^2 ),
\label{eq:Dh-expansion}
\end{eqnarray}
   where $W_{\eta\eta^\prime}$ denotes the cofactor for the element 
   $\delta_{\eta\eta^\prime} + \lambda_\eta M_{\eta\eta^\prime}$
   in $D_{\lambda_h}$, and a superscript $h$ restricts the sum to
   $\eta$ and $\eta^\prime$ corresponding to the $h$-wave channel only.
We observe, then, that close to $T_c$, Equations~(\ref{eq:gap5lin})
   factorize into two separate, independent sets of linear homogeneous
   equations for the parameters $\Delta_\eta$ representing either
   symmetries, respectively.
In the proximity of $T_c$, therefore, since $D_{\lambda_h}^0 (\mu,T) = 0$
   at $T=T_c (\mu)$, only the set of equations for $\Delta_\eta$ corresponding
   to the incipient $h$-wave channel admits a nontrivial solution,
   readily given by:
\begin{equation}
   |\Delta_\eta | = W_{\bar{\eta}\eta}^0 \epsilon,
\label{eq:homo-sol}
\end{equation}
   where $\bar{\eta} \in \{ 0,1,2 \}$, if $h=s$, 
   or $\bar{\eta} \in \{ 3,4 \}$, if $h=d$,
   and $\epsilon$ is an homogeneity factor, common for all $\eta$'s, which
   vanishes as $T\to T_c - 0$, as specified in the following.
In deriving Eq.~(\ref{eq:homo-sol}), we made use of the fact that, in the
   absence of symmetry mixing, such as at the critical point, all the 
   $\Delta_\eta$ belonging to a given symmetry channel share the same
   complex phase factor.
Inserting Eq.~(\ref{eq:gap3}) into Eq.~(\ref{eq:Dh-expansion}), and making
   use of Eq.~(\ref{eq:homo-sol}), one finds:
\begin{equation}
   \epsilon^2 = \frac{1}{\beta^3} 
   \frac{D_{\lambda_h}^0 (\mu,T)}{C_{\lambda_h}^0 (\mu,T)} ,
\label{eq:eps-0}
\end{equation}
   where
\begin{eqnarray}
   C_{\lambda_h}^0 (\mu,T) &=&
   \frac{1}{N} \sum_{\bf k}
        \frac{\phi(\beta\xi_{\bf k} /2)}{[ 1 -T_J ({\bf k}) \chi^0_{\bf k} ]^4}
   \nonumber\\
   &&{}\times
   \left( 
        {\sum_{\eta\eta^\prime}}^h 
        \lambda_\eta g_\eta ({\bf k}) W_{\eta\eta^\prime}^0 
                     g_{\eta^\prime} ({\bf k})
   \right)
   \nonumber\\
   &&\phantom{{}\times(} {}\times
   \left(
        {\sum_{\eta}}^h W_{\bar{\eta}\eta}^0 g_\eta ({\bf k})
   \right)^2 .
\label{eq:C-0}
\end{eqnarray}
We finally observe that, by construction, 
   $\lim_{T\to T_c} D_{\lambda_h}^0 (\mu,T) = 0$.
Therefore, the expansion of Eq.~(\ref{eq:eps-0}) around $T=T_c$ begins
   from the linear term in $(T-T_c )$, and one straightforwardly obtains:
\begin{equation}
   \epsilon = \alpha_h \frac{T_c}{2}
         \left( 1 - \frac{T}{T_c} \right)^{1/2} ,
\label{eq:eps-expansion}
\end{equation}
   where
\begin{mathletters}
\begin{eqnarray}
   \alpha_h^2 &=& 
   \frac{1}{C_{\lambda_h}^{0c}}
   {\sum_{\eta\eta^\prime}}^h \lambda_\eta 
         H_{\eta\eta^\prime}^{0c} W_{\eta\eta^\prime}^{0c}
         ,
   \label{eq:alphah} \\
   H_{\eta\eta^\prime}^{0} &=&
   \frac{1}{N} \sum_{\bf k} \frac{g_\eta ({\bf k}) g_{\eta^\prime} ({\bf k})}
               {[ 1 -T_J ({\bf k}) \chi_{\bf k}^0 ]^2}
               \sech^2 \left( \frac{1}{2} \beta\xi_{\bf k} \right) ,
   \label{eq:H-0}
\end{eqnarray}
\end{mathletters}
   and a superscript $c$ denotes that the limit $T\to T_c$ has been taken.
Making use of Eq.~(\ref{eq:eps-expansion}) in the expansions for 
   $|\Delta_\eta |$ and $|\Delta_{\bf k} |$, Eqs.~(\ref{eq:homo-sol}) and
   (\ref{eq:gap3}), respectively, at the critical point, one explicitly
   obtains:
\begin{mathletters}
\begin{eqnarray}
   |\Delta_\eta | &=&
      \alpha_h \frac{T_c}{2} \left( 1 - \frac{T}{T_c} \right)^{1/2} 
      W_{\bar{\eta}\eta}^{0c}  ,
\label{eq:Delta-eta-cr} \\
   |\Delta_{\bf k} | &=&
      \alpha_h \frac{T_c}{2} \left( 1 - \frac{T}{T_c} \right)^{1/2} 
      \frac{\displaystyle {\sum_\eta}^h
           W_{\bar{\eta}\eta}^{0c} g_\eta ({\bf k})}
           {\displaystyle 1-T_J ({\bf k}) \chi_{\bf k}^{0c}} .
\label{eq:Delta-k-cr}
\end{eqnarray}
\label{eq:Delta-cr}
\end{mathletters}

Equation~(\ref{eq:Delta-k-cr}) \emph{analytically} 
   yields the ${\bf k}$-dependence of the gap function at the critical
   point.
In order to exhibit more clearly the role of the ILPT amplitude 
   $T_J ({\bf k})$ in establishing such dependence, one may consider
   the limiting case in which only one basis function 
   (say, $\eta=\star$) contributes to the expansion of $\Delta_{\bf k}$.
On taking the limit $|\lambda_\star / \lambda_\eta | \to \infty$,
   $\forall\eta\neq\star$, one recovers the result
   (see also Ref.~\onlinecite{Sudboe:PhysicaC95}):
\begin{equation}
   |\Delta_{\bf k} | = \alpha_\star
   \frac{T_c}{2}\left( 1 - \frac{T}{T_c} \right)^{1/2}
   \frac{g_\star ({\bf k})}{1-T_J ({\bf k}) \chi_{\bf k}^{0c} } ,
\label{eq:Delta-cr-limit}
\end{equation}
where
\begin{equation}
   \alpha_\star^2 = \frac{\displaystyle
            \frac{1}{N} \sum_{\bf k}
            \frac{g_\star^2 ({\bf k})}{[1-T_J ({\bf k}) \chi_{\bf k}^{0c} ]^2}
            \sech^2 \left( \frac{1}{2} \beta_c \xi_{\bf k} \right)
           }{\displaystyle
            \frac{1}{N} \sum_{\bf k}
            \frac{g_\star^4 ({\bf k})}{[1-T_J ({\bf k}) \chi_{\bf k}^{0c} ]^4}
            \phi\left( \frac{1}{2} \beta_c \xi_{\bf k} \right)
           } .
\end{equation}

From Eqs.~(\ref{eq:Delta-cr}), one also recovers the critical exponent $1/2$
   analytically,
   which is typical for an order parameter at the critical point, within
   a mean-field theory.
Moreover, Equation~(\ref{eq:Delta-k-cr}) clearly shows that no symmetry
   mixing is allowed at $T=T_c$, by explicitly exhibiting which basis
   functions $g_\eta ({\bf k})$ contribute to $\Delta_{\bf k}$, and their
   weights.
The role of the ILPT mechanism is furthermore made evident by the presence
   of the factor $[1-T_J ({\bf k}) \chi_{\bf k}^{0c} ]^{-1}$ 
   in Eq.~(\ref{eq:Delta-k-cr}).
This provides the gap function $\Delta_{\bf k}$ with a remarkable anisotropy
   already at $T=T_c$, thus showing that such an anisotropy is neither
   due to self-consistency (at $T=T_c$, the values of $\Delta_{\bf k}$
   at different points in the BZ are independent of each other), nor to
   nonlinearity (at $T=T_c$, the gap equations can be linearized).
On the contrary, gap anisotropy is robust against both self-consistency and
   nonlinearity, whose relevance increases as $T$ decreases, as our numerical
   study below $T_c$ has demonstrated.

From Eq.~(\ref{eq:Delta-k-cr}) one is able to predict a line of relative
   maxima for $|\Delta_{\bf k} |$ along the $\xi_{\bf k} =0$ locus already
   at $T=T_c$.
Absolute maxima occur at the intersection of the $\xi_{\bf k} =0$ locus with
   the $\Gamma$--$X$--$M$ path, and symmetry related points.
The sharpness of the maxima is guaranteed by $T_J ({\bf k})$, and is therefore
   distinctive of the ILPT mechanism.
Away from $\xi_{\bf k} =0$, the gap function is \emph{rapidly} vanishing
   over the rest of the 1BZ, as an effect of the renormalization of the
   pair susceptibility, induced by the ILPT mechanism.
Moreover, moving along the $\xi_{\bf k} = 0$ line in ${\bf k}$-space,
   the gap function $|\Delta_{\bf k} |$ is seen to decrease \emph{more than
   linearly} as one approaches $k_x = k_y$, where $|\Delta_{\bf k} |$
   attains a minimum value, which is finite and very small, in the $s$-wave
   case, or zero, in the $d$-wave case.
This has to be contrasted with the case of a conventional $d$-wave gap,
   $\Delta_{\bf k} \propto g_3 ({\bf k})$.
In such a limit (corresponding to $T_J \to 0$ in our model), $\Delta_{\bf k}$
   would vanish \emph{linearly} as ${\bf k}$ approaches perpendicularly
   the nodal line, $k_x = k_y$.
A flat minimum (node line) along $k_x = k_y$ is indeed strongly suggested
   by ARPES results for Bi2212 single crystals,~\cite{Randeria:98a}
   and has been earlier proposed as a ``smoking gun'' for the ILPT mechanism
   by Anderson.~\cite{Anderson:battery}

The sharp anisotropic features of $|\Delta_{\bf k}|$ are robust against
   non-linearity, whose relevance increases as $T$ decreases, as shown
   by Eqs.~(\ref{eq:gap3}) and (\ref{eq:gap4}). 
Correspondingly, the normal state spectrum
   at $T=T_c$ gets gapped where $|\Delta_{\bf k}|$ is maximum \emph{far
   more significantly} than elsewhere. As $T$ decreases, one can think
   of the the Fermi line as remaining practically ungapped along disconnected
   arcs of ever smaller length. 
These arcs shrink and eventually collapse into a single point along line, 
   as $T\to 0$.

Recent ARPES experiments in underdoped
   Bi2212 single crystals by Norman {\em et al.}~\cite{Norman:97}
   are suggestive of such a scenario.
A progressive `erosion' of the Fermi line as 
   temperature decreases has been related to the opening of an unconventional
   pseudogap, precursor of the superconducting gap which opens at $T_c$.

\section{Applications}
\label{sec:applications}

In mean-field theory, the superconducting paired state is microscopically
   fully characterized by the gap function $\Delta_{\bf k}$, which we 
   now have access to over the whole 1BZ as a function of band filling and 
   temperature.
In this section, we will discuss how the solution to the gap equations 
   can be employed in the calculation of some specific thermodynamic 
   properties of the system. 
A number of physical quantities of interest have previously succesfully
   been considered with various such solutions, such as for instance the 
   anomalously large gap anisotropy
   observed in Bi2212,~\cite{Chakravarty:93a} the non-conventional features in 
   the NMR relaxation rate $1/T_1$ observed in YBCO,~\cite{Sudboe:94a}
   the variation of $T_c$ with the number of layers,~\cite{Sudboe:94b} 
   the unusual features in the neutron scattering rates observed in 
   YBCO,~\cite{Yin:97a} and a possible explanation of the spin-gap, or
   pseudo-gap.~\cite{Strong:96}
All of the above quoted calculations utilize the special features of the gap 
   that arise as a consequence of the interlayer pair-tunneling mechanism.
In particular, the calculations of the gap anisotropy, the variation of $T_c$
   with the size of the unit cell, the neutron scattering peak,
   and the spin-gap utilize the 
   unique and sharp ${\bf k}$-space features that arise in the solution to 
   the gap equations due to the unusual renormalization of the pairing 
   susceptibility, 
   $\chi_{{\bf k}} \to \chi_{{\bf k}}/[1-T_J ({\bf k}) \chi_{{\bf k}} ]$.

We choose to consider quantities that have the promise of being sensitive
   to the ${\bf k}$-space features of the gap, which are relatively readily 
   obtained, and which are  possible to confront straightforwardly with 
   experiments. 
In the following, we shall mainly focus on the specific heat anomalies of
   the model, although work is currently in progress concerning the
   in-plane coherence length and the thermal conductivity.~\cite{Angilella:98a}
These quantities are either sensitive to the presence of the particular
   $T_J$-term in the Hamiltonian such as specific heat anomalies, 
   or involve the derivative of the gap such as the coherence length and
   the thermal conductivity.
Moreover we choose, for application to the high-$T_c$ compounds, 
   parameters such 
   that the critical temperature at optimum doping is given by $T_c=90$~K. 

Although  several of the properties of the superconducting state in the 
   high-$T_c$
   compounds which in one way or another probe the ${\bf k}$-space structure 
   of the gap are unusual, the thermodynamics seems to be remarkably similar 
   to ordinary superconductors. 
This is true for instance for the entropy of the system.
Is a gap arising from an unconventional gap equation like the one considered
   in this paper, giving rise to unsual ${\bf k}$-space features in
   $\Delta_{{\bf k}}$, consistent with standard thermodynamic results otherwise
   normally associated with  conventional superconductors?~\cite{Tinkham:96}
Although not shown here, we have calculated these quantitites and found that 
   they are remarkably similar to those found in any conventional 
   superconductor.~\cite{Tinkham:96}
This is basically because quantities like entropy involve a ${\bf k}$-space
   integration over smooth functions of the gap. 
The detailed ${\bf k}$-space  features are then washed out and the results are
   to some extent quite insensitive to these features in $\Delta_{{\bf k}}$. 
The same also pertains to some extent to quantities like the NMR relaxation 
   rate $1/T_1$, 
   which exhibits features in its $T$-dependence which are reproducible
   by a gap with a number of different symmetries.~\cite{Bulut:93,Joynt:93}
This is perhaps not surprising, as this quantity involves a double
   integration over ${\bf k}$-space vectors. 
Another matter altogether is the situation where we consider
   quantities involving only one ${\bf k}$-space integration, and in 
   addition also ${\bf k}$-{\em derivatives\/} of the gap.~\cite{Angilella:98a}

\subsection{Specific heat}
\label{ssec:thermodynamics}

In this subsection, we will consider the specific heat anomalies
   of the model. 
The entropy per particle in the superconducting state is given 
   by~\cite{Leggett:75}
\begin{equation}
   S^s (\mu,T) = -2 k_B \frac{1}{N} \sum_{\bf k}
   [ f_{\bf k} \log f_{\bf k} + (1-f_{\bf k} )
   \log (1-f_{\bf k} ) ],
\label{eq:entropy}
\end{equation}
   where $f_{\bf k} = [1+\exp(\beta E_{\bf k} )]^{-1}$ is the
   Fermi function evaluated with the superconducting spectrum $E_{\bf k}$.

Differentiating $S^s (\mu,T)$, Eq.~(\ref{eq:entropy}),
   with respect to $T$ one obtains the specific heat~\cite{Leggett:75}
\begin{eqnarray}
\label{eq:cv}
   C^s_V (\mu,T) &&= T \frac{\partial S^s}{\partial T} \\
   &&= \frac{1}{2} k_B \beta^2 \frac{1}{N} \sum_{\bf k}
      E_{\bf k} \left( E_{\bf k} +  
       \beta \frac{\partial E_{\bf k}}{\partial \beta} \right)
      \sech^2 
      \left( \frac{1}{2} \beta E_{\bf k} \right). \nonumber
\end{eqnarray}
Whenever $E_{\bf k}$, i.e. $\Delta_{\bf k}$, contains discontinuities
   in its temperature-derivative as a function of $T$, the specific heat 
   Eq.~(\ref{eq:cv}) displays a finite peak. 
This is typical of the mean-field approximation, as mentioned above.
In the presence of a competition between several symmetry channels, several 
   such discontinuities may occur, at $T=T_c$ and at $T=T_m$.
However,
   we expect the height of the second peak at $T=T_m$ to be
   exponentially reduced with respect to the peak at $T=T_c$,
   due to the presence of the hyperbolic secant in Eq.~(\ref{eq:cv}).

Making use of the gap equations it is possible to derive a straightforward
   expression for $E_{\bf k} \partial E_{\bf k} /\partial\beta$, valid at
   all $T \leq T_c$, which turns out to be linear in 
   $\partial|\Delta_{\bf k} |^2 /\partial\beta$ (cf. also App.~\ref{app:mix}).
Such a quantity is numerically accessible, in principle, from the solution
   to the gap equations.
Therefore, Equation~(\ref{eq:cv}) directly yields the temperature dependence
   of $C^s_V$ also below $T_c$.
However, such dependence turns out to be conventional, and will not be shown
   here (see also Ref.~\onlinecite{Otnes:97a}).

At exactly $T=T_c$, the knowledge of the ${\bf k}$-dependence of
   $\Delta_{\bf k}$ in a closed form allows instead to study analytically
   the jump in the specific heat, normalized with respect to the specific 
   heat in the normal state, $C_V^n$, i.e. in the absence of the gap, 
   at the same temperature:
\begin{equation}
   \frac{\delta C_V^c}{C_V^{nc}} = \frac{C_V^s (\mu,T_c ) - C_V^n (\mu,T_c )}
                                        {C_V^n (\mu,T_c )} .
\label{eq:jump1}
\end{equation}
Making use of Eq.~(\ref{eq:Delta-k-cr}) corresponding to the opening of
   a generic $h$-wave symmetry gap ($h=s,d$), one readily obtains:
\begin{mathletters}
\begin{eqnarray}
\delta C_V^c &=& \frac{1}{16} k_B \alpha_h^2 \frac{1}{N} \sum_{\bf k}
   \frac{\displaystyle 
   \left( {\sum_\eta}^h W_{\bar{\eta}\eta}^{0c} g_\eta ({\bf k}) 
   \right)^2}{\displaystyle [1-T_J ({\bf k}) \chi_{\bf k}^{0c} ]^2}
   \sech^2 \left( \frac{1}{2} \beta_c \xi_{\bf k} \right) ,
\label{eq:jump2} \\
C_V^{nc} &=& 2 k_B \frac{1}{N} \sum_{\bf k}
   \left( \frac{1}{2} \beta_c \xi_{\bf k} \right)^2 
   \sech^2 \left( \frac{1}{2} \beta_c \xi_{\bf k} \right) .
\label{eq:jump3}
\end{eqnarray}
\end{mathletters}
We explicitly observe that \emph{only at $T=T_c$} one is able to include
   in Eq.~(\ref{eq:cv}) for $C_V^s$ the analytical expressions for 
   $\Delta_{\bf k}$ and its $T$-derivatives:
Numerics are only needed in performing the
   integrations over the 1BZ where required.~\cite{Lambin:84a} 
Employing the value of $T_c = T_c (\mu)$ numerically obtained as in 
   Sec.~\ref{ssec:pure} (Fig.~\ref{fig:tczero}), we are eventually able
   to evaluate the normalized jump $\delta C_V^c /C_V^{nc}$ in the specific
   heat at $T=T_c$, as a function of the chemical potential $\mu$.
We display our results in Fig.~\ref{fig:cvjump}, and compare them with
   the conventional result $\delta C_V^c /C_V^{nc} = 12/[7\zeta(3)] \simeq
   1.42613$, derived within the BCS theory for an $s$-wave, uniform
   gap function.~\cite{Enz:92,Tinkham:96}

We find a remarkable agreement with the BCS limit over an extended plateau,
   corresponding to the $s$-wave region in $\mu$.
On the contrary, a considerably lower value is obtained, on the average,
   in the $d$-wave region, including optimal doping.
On the overall, we are thus able to predict a \emph{nonuniversal} ratio
   $\delta C_V^s /C_V^{nc}$, to be contrasted with the universal BCS
   value, valid for Fermi-liquid based 
   superconductors.~\cite{Sudboe:PhysicaC95}
This is due to the widely anisotropic ${\bf k}$-dependence of the gap
   function (also close to the critical point), which is mainly traceable
   to the renormalization of the pairing susceptibility, and is thus a
   manifestation of the special nature of the interlayer pair tunneling
   mechanism.

One slightly unusual feature is the possible appearance of a \emph{second peak}
   in the specific heat at low temperatures.
Such a feature is not found in a superconductor with an order parameter
   transforming exclusively as a single basis function for an irreducible
   representation of the crystal point group $C_{4v}$.
The result we find originates from the fact that at low temperatures,
   new symmetry channels couple in to the superconducting order, as shown
   in Fig.~\ref{fig:detaT}.
This leads to a cusp in the specific heat, \emph{but not to any new diverging
   length in the problem.}
This second anomaly in the specific heat therefore \emph{does not represent
   a new superconducting phase transition,} but merely condensation of
   Cooper pairs into additional symmetry channels.
The true order parameter of the problem, $\Delta_{\bf k}$, becomes finite
   once and for all at $T=T_c$, and this point represents the only
   zero-field phase transition.
This appears to be a widely misunderstood point in the 
   literature.~\cite{Movshovich:97}
As mentioned previously, the parameters $\Delta_\eta$ do not represent
   order parameters for this problem.
Note that the second anomaly in the specific heat, at low temperatures,
   is expected to be well captured by mean-field theory.
It is located \emph{outside} the critical region of the normal
   metal-superconductor transition, while this is not the case for the
   first anomaly in the specific heat located at $T=T_c$. 
Therefore, our results for the main anomaly in the specific heat, the
   prominent step-{\em discontinuity\/} 
   at the critical point, should be replaced by a near-logarithmic 
   \emph{singularity} characteristic of the $3DXY$-model.~\cite{Nguyen:98}
This reflects the fact that for optimally doped and underdoped systems,
   phase-fluctuations in the problem appears to be strong, such that the true
   superconducting transition occurs well below the mean-field transition. 

Our main conclusion of this subsection is that in slightly overdoped
   compounds the main normalized specific heat anomaly will be mean-field 
   like, 
   but \emph{nonuniversal} due to the appearance of a renormalized pairing 
   susceptibility $\chi_{{\bf k}}/[1-T_J ({\bf k}) \chi_{{\bf k}} ]$, in 
   contrast to the standard BCS result.

\section{Summary and concluding remarks}
\label{sec:summary}

We have addressed the issue of the mixing of symmetry channels in the 
   superconducting order parameter for a bilayer superconductor 
   in the presence 
   of an interlayer pair-tunneling mechanism ~\cite{Chakravarty:93a} as 
   a possible framework for understanding numerous unconventional features 
   exhibited by the HTCS compounds.
Incipient superconductivity has been generated within each individual 
   CuO$_2$-layer through a Hubbard-like in-plane 
   potential, including primarily an on-site repulsion 
   and nearest-neighbor interaction,
   which has been strongly enhanced through the inclusion 
   of the interlayer 
   tunneling amplitude $T_J ({\bf k})$, as suggested by ARPES as well as 
   by detailed band structure calculations.

A mean-field treatment in the bilayer case allowed a computation of the
   ${\bf k}$-dependence of the in-plane order parameter 
   $\Delta_{\bf k}$.
A suitable numerical procedure has been devised in order to solve the inherently 
   nonlinear gap equations. 
It has been possible to study the evolution of the symmetry character 
   of the gap function versus temperature and chemical potential, and unveil
   a competition between $s$-wave and $d$-wave character in $\Delta_{\bf k}$.
In this description,  $\Delta_{\bf k}$ is a \emph{single complex 
   scalar} order parameter. 
No multi-component OP has to be claimed for, which would imply the existence
   of `more' condensates with different `features,' as elsewhere reported
   in the literature.~\cite{Srikanth:97a}
In particular, no transition of the normal-to-superconductor kind is
   expected when symmetries are allowed to mix: The system is already
   a superconductor with an open energy gap, whose structure in 
   ${\bf k}$-space only evolves, thus allowing pairs to condense
   into more symmetry channels.

Moreover, a surprisingly anisotropic ($s$-wave) pattern appeared to be modified 
   by the underlying symmetry character.
This is evidenced by a strongly pronounced line of maxima
   along the Fermi line, $\xi_{\bf k} =0$, which closely
   reflected the anisotropic ${\bf k}$-dependence of the interlayer
   tunneling amplitude.
Such a structure is both qualitatively and quantitatively in agreement
   with the available ARPES gap measurements in Bi2212~\cite{Ding:95}
   and recent phenomenological gap calculations starting from the multiband
   structure of the bilayer compounds.~\cite{Temmermann:96}
Moreover, it is essentially embedded in the 
   ${\bf k}$-dependence of $T_J ({\bf k})$, whereas suitable
   tuning of the intralayer coupling parameters can produce an $s$-wave
   contribution which shifts the nodes of the gap function slightly away
   from the $\Gamma$--$M$ direction ($k_x = k_y$), as reported for ARPES
   experiments in bilayer Bi2212 at a given hole content.~\cite{Ding:95} 

The gap obtained within the interlayer pair-tunneling mechanism appears to
   us to be quite promising in explaining a number of unusual properties of
   the superconducting state, such as for instance the anomalous tunneling 
   response observed in HTCS junctions.~\cite{Wei:98,Angilella:98a} 
Within the present approach, such unusal properties are associated with 
   sharp ${\bf k}$-space features of the gap due to the presence of the 
   renormalized  pair-susceptibility 
   $\chi_{{\bf k}}/[1- T_J({\bf k}) \chi_{{\bf k}}]$.

The role of the interlayer tunneling mechanism in enhancing the value of
   the critical temperature for the normal-to-superconducting
   instability,~\cite{Chakravarty:93a} as produced by a purely 
   $2D$ correlation, has been discussed and generalized for a general 
   doping level, qualitatively reproducing the universal non-monotonic 
   dependence of $T_c$ on the hole content.~\cite{Zhang:compact}

The issue of the competition in the symmetry character of the gap function 
   has been addressed both numerically and analytically  in the context of  
   the interlayer pair-tunneling mechanism.
We were able to verify that in the presence of an interlayer pair-tunneling
   matrix element, the gap symmetry is pure and cannot be mixed on an
   underlying square lattice, at the critical point. 
The gap symmetry belongs to \emph{one} of the irreducible representations of 
   $C_{4v}$, and cannot be expressed as a linear combination of several basis 
   functions of such irreducible representations. 
The one exception to this is when the chemical potential  is fine-tuned to
   a value such that accidental degeneracies occur. 
At exactly the critical point, moreover, the full ${\bf k}$-dependence and
   the critical exponent of the OP can be derived analytically, thus
   exhibiting its unconventional ${\bf k}$-space sharp structure and
   symmetry properties.

At temperatures well below $T_c$, and for  certain filling fractions,
   mixing of symetry channels may occur.
We studied the location and width of the $(\mu,T)$ region allowing
   a mixed symmetry superconducting ground state on varying the coupling
   parameters and interlayer tunneling amplitude.
In particular, we recovered prevalence of $s$-wave (resp., $d$-wave)
   symmetry at low (resp., high) band filling. 
This is due to the fact that the \emph{symmetry} of the gap is determined
   by the dominant intralayer pairing symmetry, or equivalently the dominant
   intralayer dimensionless coupling constant. 
The ``symmetry-projected" single-particle densities of states  of this problem 
   are  such that $s$-wave  coupling constant dominates at low band-fillings, 
   while $d$-wave coupling constants always dominate close to half-filling
   \cite{Otnes:97a}. 
Both $s$-wave and $d$-wave symmetries are enhanced by increasing the inter-site 
   attraction, whereas $s$-wave superconductivity is disfavored by increasing 
   the on-site repulsion.
However, the DOS argument given above also shows that the symmetry of the
  superconducting order parameter is highly dependent on doping. 

Finally, we outlined how the solution to the gap equation can be employed 
   in evaluating several quantities of interest.
In particular, we have focussed our attention to entropy and the specific
   heat anomalies of the model at the critical point.
The entropy in the superconducting state is found to have a temperature 
    variation
    very similar to  any conventional superconductor, mainly due to the fact
    that it is given by a ${\bf k}$-space integral over smooth functions
    involving the gap.
The specific heat is found to have two unusual features.
Firstly, for certain filling fractions, a mixing of symmetries may occur at a
   low temperature $T=T_m$, leading to an anomaly in the specific heat, 
   \emph{not associated with any true phase-transition.}
Secondly, there is an anomaly at the superconducting transition $T=T_c$, for 
   which our mean-field description is argued to give a reasonable description
   on the slightly overdoped side. 
This anomaly is analagous to the well-known step-discontinuity found in BCS, 
   but in our case shows a novel feature.
The normalized discontinuity turns out to be \emph{not} a universal number, 
    not only due to the different possible symmetries at the critical point,
    but also depending on the value of $T_J$, and is thus a manifestation 
    of the unusual pairing kernel in the gap equation.

\acknowledgements

Useful discussions with P. Falsaperla, J.O. Fj{\ae}restad, A.K. Nguyen, 
   J. Nyhus, and E. Otnes are gratefully acknowledged.
One of the authors (G.G.N.A.) also thanks the NTNU for warm hospitality during
   a visiting term in Trondheim, and the Istituto Nazionale di Fisica 
   Nucleare for financial support. 
One of the authors (A.S.) thanks the Norwegian Research Council for 
   financial support under Grants No.~110566/410 and 10569/410.

\appendix

\section{Numerical solution of the gap equations}
\label{app:numsolv}

For any fixed value of the chemical potential $\mu$ and temperature $T$,
   as well as of the coupling parameters $\lambda_\eta$ and interlayer 
   tunneling amplitude $T_J$, the gap parameters $\Delta_\eta$ are randomly
   initialized, and the nonlinear equation~(\ref{eq:gap3}) is solved
   for $\Delta_{\bf k}$, for each wave-vector ${\bf k}$ belonging
   to a suitably chosen fine mesh over the irreducible sector of the first
   Brillouin zone, 
   $\{ {\bf k} : 0 \leq k_x \leq \pi, 0\leq k_y \leq k_x \}$.
The values of $\Delta_{\bf k}$ thus obtained are employed to evaluate
   $M_{\eta\eta^\prime}$ through Eq.~(\ref{eq:matrices}).
Equation~(\ref{eq:gap5}) eventually defines the values of $\Delta_\eta$, to    
   be used at the successive steps in the iteration procedure.
The iterative procedure terminates when self-consistency is achieved to 
   within a preset tolerance limit in $|\Delta_{\bf k} |$.
Special care had to be used near the nodes of the gap function.
We verified the stability of the convergence procedure against the initial
   choice of $\Delta_\eta$, and 
   also by varying  the number of ${\bf k}$-points in the mesh employed in 
   the integrations.

High accuracy and a resonably small computation time in the integration
   understood in Eq.~(\ref{eq:matrices}), and elsewhere in the present paper,
   is made possible by using an adaptation to the $2D$ case of
   the analytical tetrahedron method.~\cite{Lambin:84a} 
An adaptive routine
   is suitable, due to the rapid variation of $\tilde{\chi}_{\bf k}$
   in Eq.~(\ref{eq:matrices}) for ${\bf k}$ belonging to the locus
   defined by $\xi_{\bf k} = 0$ (i.e., the Fermi surface for noninteracting
   electrons).
We carefully checked these routines by comparing the numerically evaluated DOS
   with available exact expressions.~\cite{Otnes:97a,Xing:91}

\section{Gap parameters at the mixing temperature}
\label{app:mix}

\begin{figure}
\begin{center}
\epsfig{figure=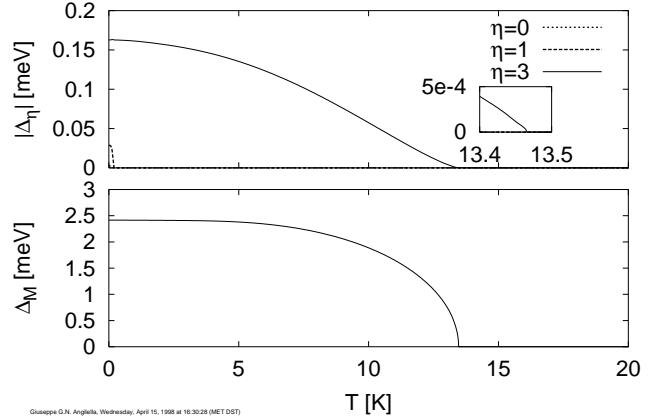,width=8.5truecm,angle=-90}
\end{center}
\caption{%
Temperature dependence of the gap parameters $|\Delta_\eta |$ (top), 
   and of the gap maximum $\Delta_M$ (bottom), at $\mu=-0.4850$~eV.
Chosen values of the in-plane coupling parameters 
   and of the interlayer tunneling amplitude 
   are $\{ \lambda_0, \lambda_1 , \lambda_2 , \lambda_3 , 
   \lambda_4 \} = \{ 
   0.01,-0.2125,0.0,-0.2125,0.0
   \}$~eV and $t_\perp = 0.08$~eV, respectively,
   yielding a critical temperature $T_c \approx 13.4$~K, 
   at which $\Delta_{\bf k}$
   opens with $d$-wave symmetry, and a mixing temperature $T_m\approx 0.2$~K,
   where $\Delta_{\bf k}$ acquires an $s$-wave contribution.
The inset in the top figure shows that $\Delta_3$ displays the expected
   critical behavior, with critical exponent $1/2$, only very close to $T_c$.
}
\label{fig:detaT}
\end{figure}

\begin{figure}
\begin{center}
\epsfig{figure=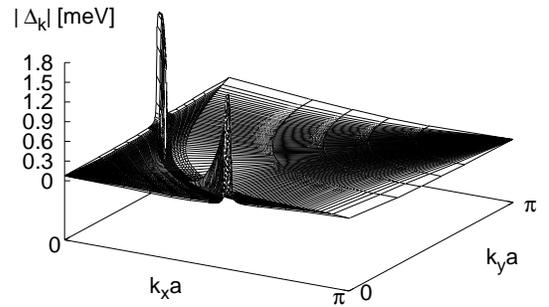,width=8.5truecm,angle=-90}
\end{center}
\caption{%
Dependence of 
   $|\Delta_{\bf k} |$ over the ${\bf k}$ in the first quarter of the
   1BZ, at $\mu=-0.4892$~eV and $T=0$.
Same values of the parameters as in Fig.~\protect\ref{fig:detaT}.
Notice the maxima structure along the $\xi_{\bf k} =0$ locus,
   including peaks at the intersection thereof with $k_y =0$ and
   $k_x =\pi/a$, and symmetry related points, whose height is enhanced
   as $T$ decreases, due to the prefactor 
   $[1-T_J ({\bf k})\chi_{\bf k} ]^{-1}$ 
   in Eq.~(\protect\ref{eq:gap2}). 
}
\label{fig:dkT}
\end{figure}

\begin{figure}
\begin{center}
\epsfig{figure=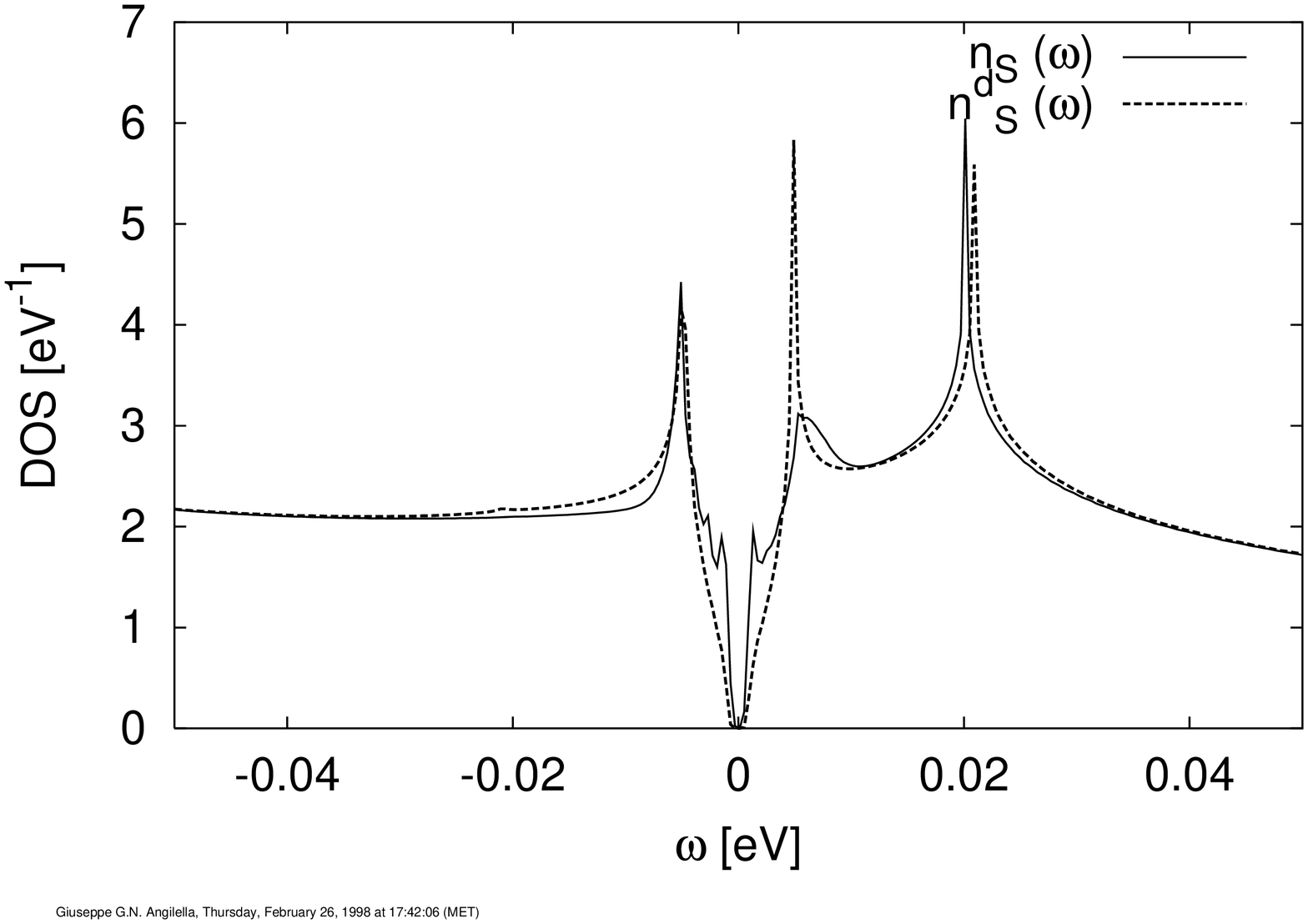,width=8.5truecm,angle=-90}
\end{center}
\caption{%
Superconducting DOS $n_S (\omega)$, corresponding to
   an anisotropic ${\bf k}$-dependent gap in the presence of ILPT
   (continuous line), and $n_S^d (\omega)$, corresponding to a purely $d$-wave
   gap, without ILPT (dashed line), at $\mu=-0.47$~eV, $T=0$~K.
Same values of the parameters as in Fig.~\protect\ref{fig:detaT}. 
}
\label{fig:sDOS}
\end{figure}

\begin{figure}
\begin{center}
\epsfig{figure=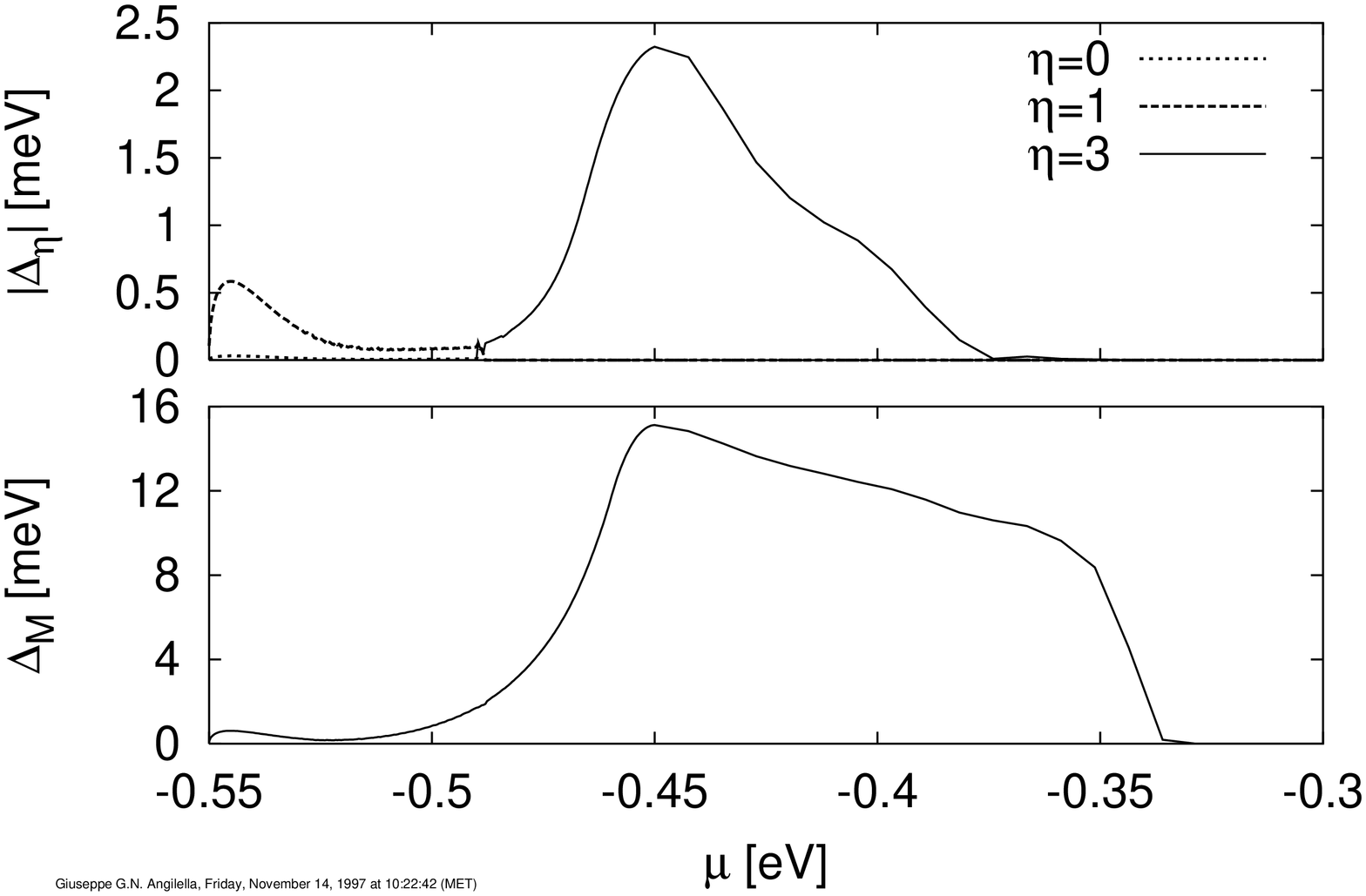,width=8.5truecm,angle=-90}
\end{center}
\caption{%
Dependence of the gap parameters $|\Delta_\eta (T=0)|$ and of the
   gap maximum at zero temperature $\Delta_M$ on the chemical potential
   $\mu$.
Same values of the parameters as in Fig.~\protect\ref{fig:detaT}. 
}
\label{fig:detamu}
\end{figure}

\begin{figure}
\begin{center}
\epsfig{figure=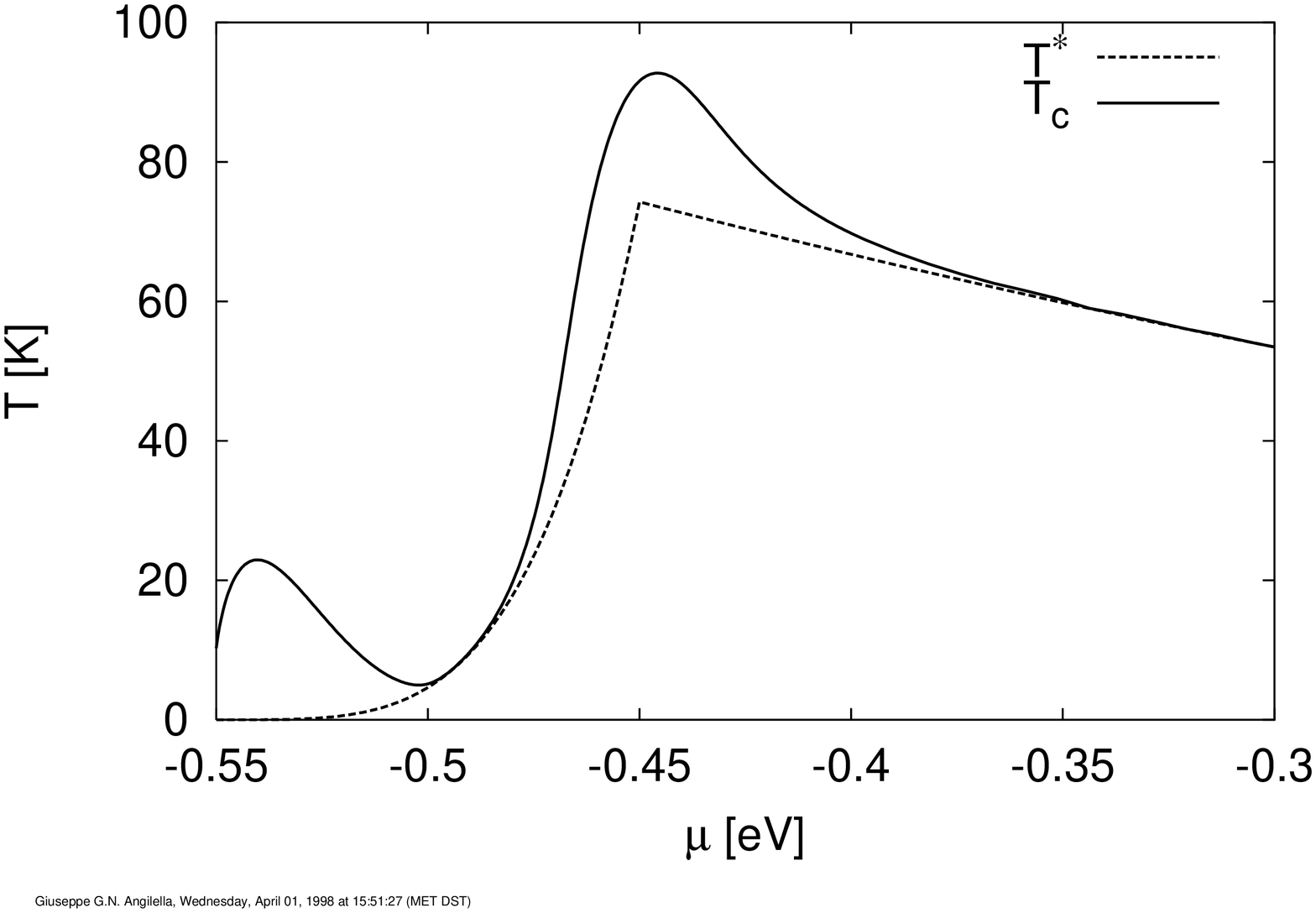,width=8.5truecm,angle=-90}
\end{center}
\caption{%
Lower bound temperature $T^\star (\mu)$ (dashed line) and critical 
   temperature $T_c (\mu)$ (continous line), as functions of the
   chemical potential $\mu$.
Same values of the parameters as in Fig.~\protect\ref{fig:detaT}. 
}
\label{fig:tczero}
\end{figure}

\begin{figure}
\begin{center}
\epsfig{figure=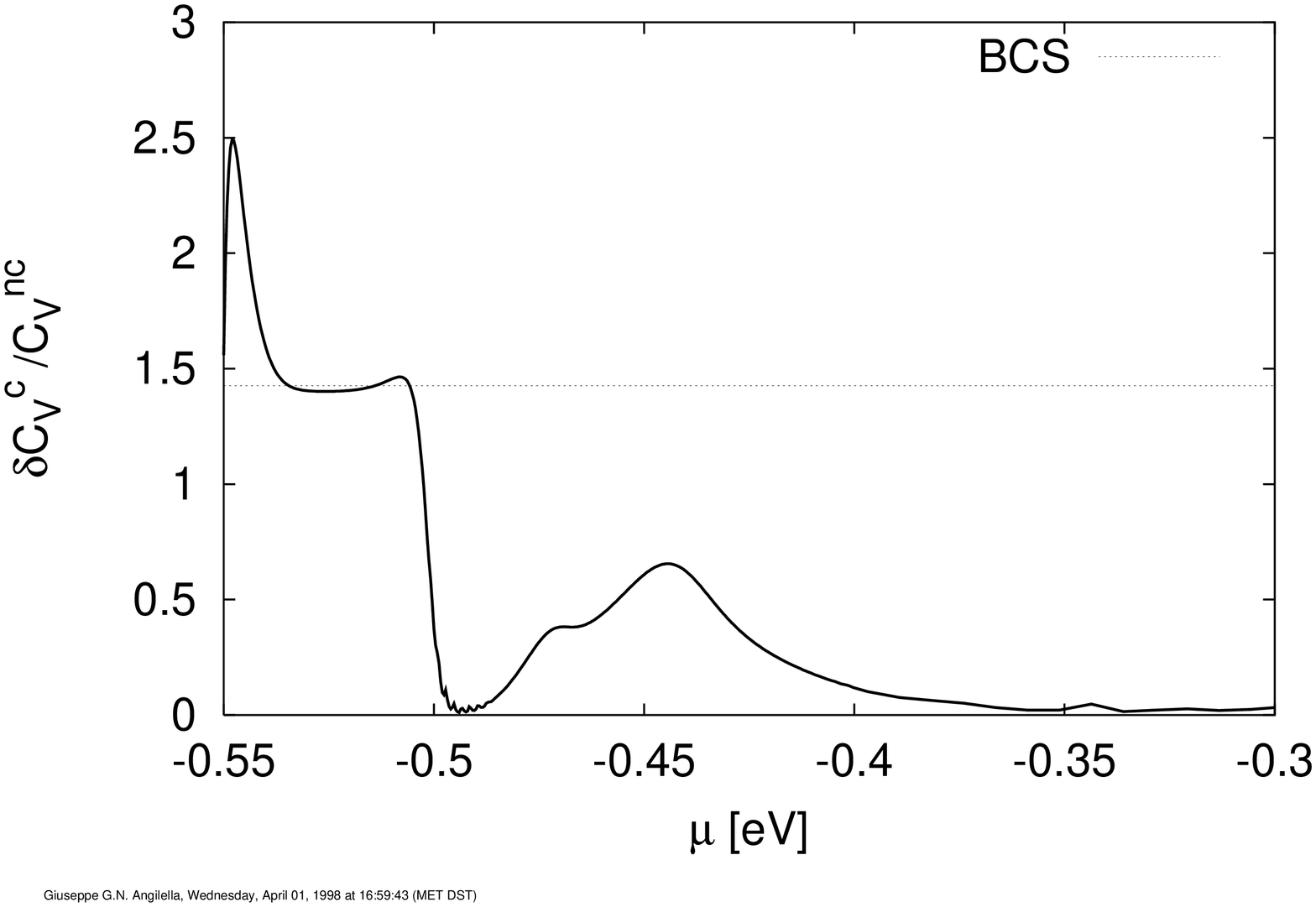,width=8.5truecm,angle=-90}
\end{center}
\caption{%
Normalized jump $\delta C_V^c /C_V^{nc}$ in the specific heat at $T=T_c$
   within the ILPT mechanism, as a function of the chemical potential
   $\mu$ (continuous line). 
Like in Fig.~\protect\ref{fig:detaT}, 
   we used $\{ \lambda_0, \lambda_1 , \lambda_2 , \lambda_3 , \lambda_4 \} 
   = \{ 0.01,-0.2125,0.0,-0.2125,0.0 \}$~eV 
   and $t_\perp = 0.08$~eV, respectively,
The BCS universal limit $12/[7\zeta(3)] \simeq
   1.42613$ is also shown, for comparison (dashed line).
}
\label{fig:cvjump}
\end{figure}

Following a procedure analogous to that outlined in Sec.~\ref{ssec:anisTc},
   one is able to derive a critical exponent $1/2$ also for the gap components
   $\Delta_\eta$ which open at the mixing temperature $T=T_m$, thus endowing
   the gap function with an additional contribution with a generic $h$-wave
   symmetry, orthogonal to the one already present.
Such a result is consistent with the conventional case ($T_J \to 0$), and
   is illustrated by the numerical example shown in Fig.~\ref{fig:detaT}.
The calculations are more involved, although straightforward, and will not
   be shown here in detail.
They must however take into account that a gap is \emph{already} open 
   at $T=T_m$.
Generalizing the notation introduced in Sec.~\ref{ssec:anisTc}, the final
   result is:
\begin{equation}
   |\Delta_\eta | = \gamma_h \frac{T_m}{2} \left( 1 - \frac{T}{T_m} 
                    \right)^{1/2} W_{\bar{\eta}\eta}^m ,
\end{equation}
   where
\begin{equation}
   \gamma_h^2 =
   \frac{1}{C_{\lambda_h}^m}
   {\sum_{\eta\eta^\prime}}^h \lambda_\eta 
         H_{\eta\eta^\prime}^m W_{\eta\eta^\prime}^m ,
\end{equation}
   and
\begin{eqnarray}
C_{\lambda_h}^m &=& \frac{1}{N} \sum_{\bf k}
   \frac{\phi(\beta_m E_{\bf k}^m /2)}{[1-T_J ({\bf k})\chi_{\bf k}^m ]^3}
   \nonumber\\
   &&{}\times
   \left\{ 1 -T_J ({\bf k}) \left[\chi_{\bf k}^m -2\beta_m^3 \phi\left( 
   \frac{1}{2}
   \beta_m E_{\bf k}^m \right) |\Delta_{\bf k} |^2 \right] \right\}^{-1}
   \nonumber\\
   &&\phantom{{}\times} \times
   \left( {\sum_{\eta\eta^\prime}}^h \lambda_\eta g_\eta ({\bf k})
   W_{\eta\eta^\prime}^m g_{\eta^\prime} ({\bf k}) \right)
   \nonumber\\
   &&\phantom{{}\times\times} \times
   \left( {\sum_\eta}^h W_{\bar{\eta}\eta}^m g_\eta ({\bf k}) \right)^2 ,\\
H_{\eta\eta^\prime}^m &=& \frac{1}{N} \sum_{\bf k}
   \frac{g_\eta ({\bf k}) g_{\eta^\prime} ({\bf k})}%
   {[1-T_J ({\bf k})\chi_{\bf k}^m ]^2}
   \left[ \sech^2 \left( \frac{1}{2} \beta_m E_{\bf k}^m \right) \right.
   \nonumber\\
   &&{}+
   \left. 4\beta_m \phi \left( \frac{1}{2} \beta_m E_{\bf k}^m \right)
   \left( \frac{\partial |\Delta_{\bf k} |^2}{\partial T} \right)_{T=T_m +0}
   \right].
\end{eqnarray}
An index $m$ denotes the inclusion of the gap
   function without the new $h$-wave contribution, and that the limit
   $T\to T_m$ has been taken afterwards.

The above equations reduce to the analogous ones derived in 
   Sec.~\ref{ssec:anisTc} at $T=T_c$ and $\Delta_{\bf k} =0$.
The latter expressions are analytical but not in closed form, 
   since they require
   the knowledge of $\beta(\partial |\Delta_{\bf k} |^2 /\partial T)$, 
   at $T=T_m +0$, which can be accessed only numerically.
However, one expects such quantity to be vanishingly small
   for $T_m \ll T_c$, like in the conventional case.
(It would be exactly zero if $T_m = 0$).

\bibliography{main,notes,compact,new,newer}

\begin{thebibliography}{10}
\bibitem{note:7}
On the other hand, it may be worth emphasizing that the physical origin of
  in-plane pairing is logically independent from the ILPT mechanism, as earlier
  recognized by Anderson (see, e.g., Ref.~\protect\onlinecite{Anderson:97}).
\bibitem{note:6}
It is understood that normal Hartree terms, i.e. of the form $\propto c_{{\bf
  k}\uparrow}^{i\dag} c_{{\bf k}^\prime \uparrow}^j \langle c_{{\bf
  -k}\downarrow}^{i\dag} c_{{\bf -k^\prime}\downarrow}^j \rangle$, can be
  absorbed in the renormalization of the chemical potential, which is here
  treated as a free parameter.
\bibitem{note:2}
Note that it is the presence of a generally nonzero $\Delta_{\bf k}$ in the
  pair susceptibility $\chi_{\bf k}$ which prevents the quantity $1-T_J ({\bf
  k}) \chi_{\bf k}$ from being actually zero for $T<T_c$.
\end{thebibliography}
\bibliographystyle{mprsty}

\vspace{12truecm}

\end{document}